\documentclass[preprint,aps,preprintnumbers,nofootinbib,superscriptaddress,showpacs]{revtex4}
\usepackage{amsmath,amssymb,amsbsy}
\usepackage{graphicx}
\usepackage{epsfig}
\usepackage{bm}
\usepackage{psfrag}
\usepackage{color}
\usepackage{rotating}

\usepackage{url}
\newenvironment{CFigure}[1][tbp]{\begin{figure}[#1]\centering}%
                                {\end{figure}}

\newenvironment{CTable}[1][tbp]{\begin{table}[#1]\centering}%
                                {\end{table}}

\newcommand*{\ie}{\textit{i.e.},\ }

\def\etal{{\it et al.}}

\def\CO{{\cal O}}

\def\CV{{\cal V}}

\def\bar{\overline}
\def\hat{\widehat}
\def\tilde{\widetilde}

\def\spose#1{\hbox to 0pt{#1\hss}}
\def\ltapprox{\mathrel{\spose{\lower 3pt\hbox{$\mathchar"218$}}
 \raise 2.0pt\hbox{$\mathchar"13C$}}}
\def\gtapprox{\mathrel{\spose{\lower 3pt\hbox{$\mathchar"218$}}
 \raise 2.0pt\hbox{$\mathchar"13E$}}}
\def\inapprox{\mathrel{\spose{\lower 3pt\hbox{$\mathchar"218$}}
 \raise 2.0pt\hbox{$\mathchar"232$}}}

\begin{document}

\vphantom{}

\title{\boldmath
The $B\to\pi\ell\nu$ semileptonic form factor from three-flavor lattice QCD:  A model-independent determination of $|V_{ub}|$}

\author{Jon~A.~Bailey}
\affiliation{Fermi National Accelerator Laboratory, Batavia, Illinois, USA}

\author{C.~Bernard}
\affiliation{Department of Physics, Washington University, St.~Louis, Missouri, USA}

\author{C.~DeTar}
\affiliation{Physics Department, University of Utah, Salt Lake City, Utah, USA}

\author{M.~Di Pierro}
\affiliation{School of Computer Sci., Telecom.\ and Info.\ Systems, DePaul University, Chicago, Illinois, USA}

\author{A.X.~El-Khadra}
\affiliation{Physics Department, University of Illinois, Urbana, Illinois, USA}

\author{R.T.~Evans}
\affiliation{Physics Department, University of Illinois, Urbana, Illinois, USA}

\author{E.~D.~Freeland}
\affiliation{Liberal Arts Department, The School of the Art Institute of Chicago, Chicago, Illinois, USA }

\author{E.~Gamiz}
\affiliation{Physics Department, University of Illinois, Urbana, Illinois, USA}

\author{Steven~Gottlieb}
\affiliation{Department of Physics, Indiana University, Bloomington, Indiana, USA}

\author{U.M.~Heller}
\affiliation{American Physical Society, One Research Road, Ridge, New York, USA}

\author{J.E.~Hetrick}
\affiliation{Physics Department, University of the Pacific, Stockton, California, USA}

\author{A.S.~Kronfeld}
\affiliation{Fermi National Accelerator Laboratory, Batavia, Illinois, USA}

\author{J.~Laiho}
\affiliation{Department of Physics, Washington University, St.~Louis, Missouri, USA}

\author{L.~Levkova}
\affiliation{Physics Department, University of Utah, Salt Lake City, Utah, USA}

\author{P.B.~Mackenzie}
\affiliation{Fermi National Accelerator Laboratory, Batavia, Illinois, USA}

\author{M.~Okamoto}
\affiliation{Fermi National Accelerator Laboratory, Batavia, Illinois, USA}

\author{J.~N.~Simone}
\affiliation{Fermi National Accelerator Laboratory, Batavia, Illinois, USA}

\author{R.~Sugar}
\affiliation{Department of Physics, University of California, Santa Barbara, California, USA}

\author{D.~Toussaint}
\affiliation{Department of Physics, University of Arizona, Tucson, Arizona, USA}

\author{R.S.~Van~de~Water}
\email[]{ruthv@bnl.gov}
\affiliation{Fermi National Accelerator Laboratory, Batavia, Illinois, USA}

\collaboration{Fermilab Lattice and MILC Collaborations}
\noaffiliation

\date{\today}

\begin{abstract}
We calculate the form factor $f_+(q^2)$ for $B$-meson semileptonic decay in unquenched lattice QCD with 2+1 flavors of light sea quarks.  We use Asqtad-improved staggered light quarks and a Fermilab bottom quark on gauge configurations generated by the MILC Collaboration.  We simulate with several light quark masses and at two lattice spacings, and extrapolate to the physical quark mass and continuum limit using heavy-light meson staggered chiral perturbation theory.  We then fit the lattice result for $f_+(q^2)$ simultaneously with that measured by the BABAR experiment using a parameterization of the form factor shape in $q^2$ which relies only on analyticity and unitarity in order to determine the CKM matrix element $|V_{ub}|$.  This approach reduces the total uncertainty in $|V_{ub}|$ by combining the lattice and experimental information in an optimal, model-independent manner.  We find a value of $|V_{ub}| \times 10^3 = 3.38 \pm 0.36$.  
\end{abstract}

\pacs{12.15.Hh, 12.38.Gc, 13.20.He}
\maketitle

\section{Introduction}
\label{sec:Intro}

The semileptonic decay $B \to \pi \ell \nu$ is a sensitive probe of the
heavy-to-light quark-flavor changing $b\to u$ transition.  When
combined with an experimental measurement of the differential decay
rate, a precise QCD determination of the $B \to \pi \ell \nu$ form
factor allows a clean determination of the Cabibbo-Kobayashi-Maskawa
(CKM) matrix element $|V_{ub}|$ with all sources of systematic
uncertainty under control.  In the Standard Model, the differential decay rate for this process is
\begin{equation}
	\frac{d\Gamma(B\to\pi\ell\nu)}{dq^2} = \frac{G_F^2 |V_{ub}|^2}{192 \pi^3 m_B^3} \left[
(m_B^2 + m_\pi^2 - q^2)^2 - 4 m_B^2 m_\pi^2 \right]^{3/2} |f_+(q^2)|^2 ,
\end{equation}
where $q \equiv p_B - p_\pi$ is the momentum transferred from the
$B$-meson to the outgoing leptons.  The form factor, $f_+(q^2)$,
parameterizes the hadronic contribution to the weak decay, and must be
calculated nonperturbatively from first principles using lattice QCD.  

A precise knowledge of CKM matrix elements such as $|V_{ub}|$ is
important not only because they are fundamental parameters of the
Standard Model, but because inconsistencies between independent
determinations of the CKM matrix elements and $CP$-violating phase
would provide evidence for new physics.  Although the Standard Model
has been amazingly successful in describing the outcome of  most
particle physics experiments to date, it cannot account for gravity, dark matter and dark energy, or the large matter-antimatter asymmetry of the universe.
Thus we know that it is incomplete, and expect new physics to affect
the quark-flavor sector to some degree, although we do not know \emph{a
priori} what experimental and theoretical precision will be needed to
observe it.  

The determination of $|V_{ub}|$ from $B \to \pi \ell \nu$ semileptonic
decay relies upon the assumption that, because the leading Standard
Model decay amplitude is mediated by tree-level $W$-boson exchange, it
will not be significantly affected by new physics at the current level
of achievable precision.  Recently, however, hints of new physics have
appeared in various regions of the heavy-quark flavor sector such as
$CP$-asymmetries in $B\to K \pi$~\cite{BelleNature}, constraints on
$\textrm{sin}(2\beta)$ from $\Delta F=2$ neutral meson mixing and
1-loop penguin-induced decays~\cite{Lunghi:2008aa}, and the phase of
the $B_s$-mixing amplitude~\cite{Aaltonen:2007he,:2008fj,Bona:2008jn}.  The unexpected
inconsistency most relevant to our new lattice QCD calculation of the
$B\to \pi \ell \nu$ form factor and $|V_{ub}|$ is the current
``$f_{D_s}$ puzzle"~\cite{Dobrescu:2008er}.  The HPQCD Collaboration's
lattice QCD calculation of the $D_s$-meson leptonic decay
constant $f_{D_s}$~\cite{Follana:2007uv} disagrees with the latest
results from the Belle, BABAR, and CLEO
experiments~\cite{:2007ws,Aubert:2006sd,Pedlar:2007za,:2007zm,Stone:2008gw} at the
3-$\sigma$ level, although HPQCD's determinations of the masses $m_{D^+}$ and $m_{D_s}$ and the decay constants $f_\pi$, $f_K$,
and $f_{D^+}$ all agree quite well with experimental
measurements~\cite{Amsler:2008zz,:2008sq}.  Furthermore, because the significance of the discrepancy is dominated by the statistical experimental uncertainties, it cannot easily be
explained by an underestimate of the theoretical uncertainties.
Additional lattice QCD calculations of $f_{D_s}$ are needed to either
confirm or reduce the inconsistency.  If the disagreement holds up,
however, it is evidence for a large new physics contribution to a
tree-level Standard Model process at the few percent-level.  Therefore,
although $B\to \pi \ell \nu$ semileptonic decay provides a
theoretically clean method for determining $|V_{ub}|$ within the
framework of the Standard Model, we should keep in mind that new
physics could appear in $b\to u$ transitions.

\bigskip

Understanding and controlling all sources of systematic uncertainty in
lattice QCD calculations of hadronic weak matrix elements is essential
in order to allow accurate determinations of Standard Model parameters
and reliable searches for new physics.  The hadronic amplitudes for
$B\to \pi \ell\nu$, in particular, can be calculated accurately using
current lattice QCD methods because the decay process is ``gold
plated", i.e., there is only a single stable hadron in both the initial
and final states.  Lattice calculations with staggered quarks allow for
realistic QCD simulations with dynamical quarks as light as $m_s/10$,
multiple lattice spacings, large physical volumes, and high statistics.
The resulting simulations of many light-light and heavy-light meson
quantities with dynamical staggered quarks are in excellent numerical
agreement with experimental results \cite{Davies:2003ik}.  This
includes both postdictions, such as the pion decay constant
\cite{Aubin:2004fs}, and predictions, as in the case of the $B_c$
meson mass \cite{Allison:2004be}.  Such successes show that the
systematic uncertainties in these lattice QCD calculations are under
control, and give confidence that additional calculations using the
same methods are reliable.  

The publicly available MILC gauge configurations with three flavors of
improved staggered quarks~\cite{Bernard:2001av} that have enabled
these precise lattice calculations make use of the ``fourth-root"
procedure for removing the undesired four-fold degeneracy of staggered
lattice fermions.  Although this procedure has not been rigorously proven correct,
Shamir uses plausible assumptions to argue that the continuum limit of the rooted theory is in the same
universality class as QCD~\cite{Shamir:2004zc,Shamir:2006nj}.   The rooting procedure leads to
violations of unitarity that vanish in the continuum limit; both
theoretical arguments~\cite{Bernard:2006zw,Bernard:2007ma} and
numerical
simulations~\cite{Prelovsek:2005rf,Bernard:2007qf,Aubin:2008wk}, however, show that the unitarity-violating lattice artifacts in the pseudo-Goldstone boson sector can be
described and hence removed using rooted Staggered Chiral Perturbation
Theory (rS$\chi$PT), the low-energy effective description of the rooted
staggered lattice theory~\cite{Lee:1999zx,Aubin:2003mg,Sharpe:2004is}.
Given the wealth of numerical and analytical evidence supporting the
validity of the rooting procedure, most of which is reviewed in
Refs.~\cite{Durr:2005ax,SharpePlenary,Kronfeld:2007ek}, we work under
the plausible assumption that the continuum limit of the rooted
staggered theory is QCD.  We note, however, that it is important to
have crosschecks of lattice calculations of
phenomenologically-important quantities using a variety of fermion
formulations, since they all have different sources of systematic
uncertainty.

Both existing unquenched lattice calculations of the $B\to\pi\ell\nu$
form factor use the MILC configurations.  When combined with the Heavy
Flavor Averaging Group's latest determination of the experimental decay
rate from ICHEP 2008~\cite{HFAG_ICHEP08}, they yield the following values for
$|V_{ub}|$:
\begin{eqnarray}
|V_{ub}| \times 10^3 & = & 3.40 \pm 0.20 ^{+0.59}_{-0.39} \qquad\qquad
\text{HPQCD \cite{Dalgic:2006dt}}, \\
|V_{ub}| \times 10^3 & = & 3.62 \pm 0.22 ^{+0.63}_{-0.41}  \qquad\qquad
\text{Fermilab-MILC \cite{Okamoto:2004xg}} ,
\end{eqnarray}
where the errors are experimental and theoretical, respectively.  
Both analyses primarily rely upon data generated at a ``coarse" lattice spacing of $a \approx 0.12$ fm, and use a smaller amount of ``fine" data at $a \approx 0.09$ fm to check the estimate of discretization errors.  Neither is therefore able to extrapolate the $B\to\pi\ell\nu$ form factor to the continuum ($a \to 0$).  The most significant difference in the two calculations is their use of different lattice formulations for the bottom quarks.  The HPQCD Collaboration~\cite{Dalgic:2006dt} uses nonrelativistic (NRQCD) heavy quarks~\cite{Lepage:1992tx}, whereas we use relativistic clover quarks with the Fermilab interpretation~\cite{ElKhadra:1996mp} via heavy quark effective theory (HQET)~\cite{Kronfeld:2000ck,Harada:2001fi,Harada:2001fj}.  Both methods work quite well for heavy
bottom quarks.  The Fermilab treatment, however, has the advantage that
it can also be applied to charm quarks;  we can therefore use the same
method for other semileptonic form factors such as $D\to\pi \ell \nu$, $D \to K \ell \nu$, and $B\to D^*\ell\nu$~\cite{Aubin:2005ar,Bernard:2008dn}.  The two unquenched lattice
calculations of the $B\to\pi\ell\nu$ form factor, which have largely
independent sources of systematic uncertainty, nevertheless lead to
consistent values of $|V_{ub}|$ with similar total errors of $\sim 15\%$.

\bigskip

In this paper we present a new \emph{model-independent} unquenched
lattice QCD calculation of the $B\to\pi\ell\nu$ semileptonic form
factor and $|V_{ub}|$.  Our work builds upon the previous Fermilab-MILC
calculation and improves upon it in several ways.  We now include data
on both the coarse and fine MILC lattices, and can therefore take the
$a\to 0$ limit of our data which is generated at nonzero lattice
spacing.  We also have additional statistics on a subset of the coarse
ensembles.  The most important improvements, however, are in the
analysis procedures.

We have removed all model-dependent assumptions about the shape in
$q^2$ of the form factor from the current analysis.  Our result is
therefore theoretically cleaner and more reliable than those of
previous lattice QCD calculations.  The first refinement over previous
unquenched lattice $B\to\pi\ell\nu$ form factor calculations is in the
treatment of the chiral and continuum extrapolations.   We
simultaneously extrapolate to physical quark masses and zero lattice
spacing and interpolate in the momentum transfer $q^2$ by performing a
single fit to our entire data set (all values of $m_q$, $a$, and $q^2$)
guided by functional forms derived in heavy-light meson staggered
chiral perturbation theory (HMS$\chi$PT)~\cite{Aubin:2007mc}.  We
thereby extract the physical form factor $f_+(q^2)$ in a controlled
manner without introducing a particular ansatz for the form factor's
$q^2$ dependence.  The second refinement over previous unquenched
$B\to\pi\ell\nu$ lattice form factor calculations is in the combination
of the lattice form factor result and experimental data for the decay
rate to determine the CKM matrix element $|V_{ub}|$.  We fit our
lattice numerical Monte Carlo data and the 12-bin BABAR experimental
data~\cite{Aubert:2006px} together to the model-independent
``$z$-expansion" of the form factor given in
Ref.~\cite{Arnesen:2005ez}, in which the form factor is described by a
power series in a small quantity $z$ with the sum of the squares of the series
coefficients bounded by unitarity constraints.  We leave the relative
normalization factor, $|V_{ub}|$, as a free parameter to be determined
by the fit, thereby extracting $|V_{ub}|$ in an optimal,
model-independent way.  Others have also fit lattice and experimental
results together using different, equally-valid,
parameterizations~\cite{Flynn:2007ii,Bourrely:2008za}.  This work,
however, is the first to use the full correlation matrices, derived
directly from the data, for both the lattice calculation and
experimental measurement. 

\bigskip

This paper is organized as follows.  In Sec.~\ref{sec:SimDetails} we
describe the details of our numerical simulations. We discuss the
gluon, light-quark, and heavy-quark lattice actions, and present the
parameters used, such as the quark masses and lattice spacings.  We
then define the matrix elements needed to calculate the semileptonic
form factors and discuss the method for matching the lattice
heavy-light current to the continuum.   Next we describe our analysis
for determining the form factors in Sec.~\ref{sec:Analysis}.  This is a
three-step procedure.   We first fit pion and $B$-meson 2-point
correlation functions to extract the meson masses.  We then fit the
$B \to \pi$ 3-point function, using the masses and
amplitudes from the 2-point fits as input, to extract the lattice form
factors at each value of the light quark mass and lattice spacing.
Finally, we extrapolate the results at unphysical quark masses and
nonzero lattice spacing to the physical light quark masses and zero
lattice spacing using HMS$\chi$PT.  In Sec.~\ref{sec:Errors} we estimate
the contributions of the various systematic uncertainties to the form
factors, discussing each item in the error budget separately.  We then
present the final result for $f_+(q^2)$ with a detailed breakdown of
the error by source in each $q^2$ bin.  We combine our result for the
form factor with experimental data from the BABAR Collaboration to
determine the CKM matrix element $|V_{ub}|$ in Sec.~\ref{sec:Vub}.  We
also define the model-independent description of the form factor shape
that we use in the fit and discuss alternative parameterizations of the
form factor.  Finally, in Sec.~\ref{sec:Conc} we compare our results with
those of previous unquenched lattice calculations.  We also compare our
determination of $|V_{ub}|$ with inclusive determinations and to the
preferred values from the global CKM unitarity triangle analysis.
We conclude by discussing the prospects for
improvements in our calculation and its impact on searches
for new physics in the quark flavor sector.

\section{Lattice Calculation}
\label{sec:SimDetails}

In this section we describe the details of our numerical lattice simulations.  We first present the actions and parameters used for the light (up, down, strange) and heavy (bottom) quarks in Sec.~\ref{sec:Params}.  We then define the procedure for constructing lattice correlation functions with both staggered light and Wilson heavy quarks in Sec.~\ref{sec:HLCorrs}.  Finally, in Sec.~\ref{sec:HLCurrents}, we show how to match the lattice heavy-light vector currents to the continuum with a mostly nonperturbative method, so that lattice perturbation theory is only needed to estimate a small correction.  

\subsection{Actions and Parameters}
\label{sec:Params}

We use the ensembles of lattice gauge fields generated by the MILC Collaboration and described in Ref.~\cite{Bernard:2001av} at two lattice spacings, $a \approx 0.12$ and $0.09$~fm, in our numerical lattice simulations of the $B\to\pi\ell\nu$ semileptonic form factor.   The ensembles include the effects of three dynamical staggered quarks --- two degenerate light quarks with masses ranging from $m_s/8$ -- $m_s/2$ and one heavier quark tuned to within 10--30\% of the physical strange quark mass.  This allows us to perform a controlled extrapolation to both the continuum and the physical average $u$-$d$ quark mass.  The physical lattice volumes are all sufficiently large ($m_\pi L \gtapprox 4$) to ensure that effects due to the finite spatial extent remain small.

For each independent ensemble we compute the light valence quark in the 2-point and 3-point correlation functions only at the same mass, $m_l$, as the light quark in the sea sector.  Thus all of our simulations are at the ``full QCD" point.   Note, however, that we still have many correlated data points on each ensemble because of the multiple pion energies.  Table~\ref{tab:MILC_ens} shows the combinations of lattice spacings, lattice volumes, and quark masses used in our calculation.

\begin{CTable}
\caption{Lattice simulation parameters.  The columns from left to right are the approximate lattice spacing in fm, the bare light quark masses $am_l/am_s$, the linear spatial dimension of the lattice in fm, the dimensionless factor $m_\pi L$ (corresponding to the taste-pseudoscalar pion composed of light sea quarks), the dimensions of the lattice in lattice units, the number of configurations used for this analysis, the clover term $c_{SW}$ and bare $\kappa$ value used to generate the bottom quark, and the improvement coefficient used to rotate the bottom quark field in the $b\to u$ vector current.\medskip}
\label{tab:MILC_ens}

\begin{tabular}{cccccccccc}

  \hline \hline
  $a$(fm) & $am_l/ am_s$ & $L$(fm) & \ $m_\pi L$ \ & Volume & \ $\#$ Configs. & $c_{SW}$ & $\kappa_b$ & $d_1$ \\ 
  
   \hline
  $0.09$ & $0.0062/0.031$ & 2.4 & 4.1 & $28^3\times 96$ & 557  & \ 1.476 & 0.0923 & \ 0.09474\\
  $0.09$ & $0.0124/0.031$ & 2.4 & 5.8 & $28^3\times 96$ & 518  & \ 1.476 & 0.0923 & \ 0.09469\\

  \hline
   $0.12$ & $0.005/0.05$ & 2.9 & 3.8 & $24^3\times 64$ & 529 & \ 1.72 & 0.086 & \ 0.09372 \\
   $0.12$ & $0.007/0.05$ & 2.4 & 3.8 & $20^3\times 64$ & 836 & \ 1.72 & 0.086 & \ 0.09372 \\
   $0.12$ & $0.01/0.05$ & 2.4 & 4.5 & $20^3\times 64$ & 592 & \ 1.72 & 0.086 & \ 0.09384 \\
   $0.12$ & $0.02/0.05$ & 2.4 & 6.2 & $20^3\times 64$ & 460 & \ 1.72& 0.086 & \ 0.09368 \\

\hline\hline
  
\end{tabular}\end{CTable}

For bottom quarks in 2-point and 3-point correlation functions we use the Sheikholeslami-Wohlert (SW) ``clover'' action~\cite{Sheikholeslami:1985ij} with the Fermilab interpretation via HQET~\cite{ElKhadra:1996mp,Kronfeld:2000ck}, which is well-suited  for heavy quarks, even when $a m_Q \gtapprox 1$.  Because the spin-flavor symmetry of heavy quark systems is respected by the lattice regulator, the expansion in  $1/m_Q$ of the heavy-quark lattice action has the same form as the $1/m_Q$ expansion of the heavy-quark part of the QCD action.  Discretization effects in the lattice heavy-quark action are therefore parameterized order-by-order in the heavy-quark expansion by deviations of effective operator coefficients from their values in continuum QCD.  Thus, in principle, the lattice heavy-quark action can be improved to arbitrarily high orders in HQET by tuning a sufficiently large number of parameters in the lattice action.  In practice, we tune the hopping parameter, $\kappa$, and the clover coefficient, $c_{\mathrm{SW}}$, of the SW action, to remove discretization effects through next-to-leading order, $\CO(1/m_Q)$, in the heavy-quark expansion.

The SW action includes a dimension-five interaction with a coupling $c_{\textrm{SW}}$ that must be adjusted to normalize the heavy quark's chromomagnetic moment correctly~\cite{ElKhadra:1996mp}.  In our calculation we set the value of $c_{\textrm{SW}} = u_0^{-3}$, as suggested by tadpole-improved, tree-level perturbation theory~\cite{Lepage:1992xa}. We determine the value of $u_0$ either from the plaquette ($a\approx 0.09$ fm) or from the Landau link ($a \approx 0.12$ fm).  The tadpole-improved bare quark mass for SW quarks is given by
\begin{eqnarray} 
am_0 = \frac{1}{u_0}\left(\frac{1}{2\kappa}-\frac{1}{2\kappa_{\rm crit}}\right),
\end{eqnarray}
such that tuning the parameter $\kappa$ to the critical quark hopping parameter $\kappa_{\rm crit}$ leads to a massless pion.  Before generating the correlation functions needed for the $B\to\pi\ell\nu$ form factor, we compute the spin-averaged $B_s$ kinetic mass on a subset of the available ensembles in order to tune the bare $\kappa$ value for bottom (and hence the corresponding bare quark mass) to its physical value~\cite{ElKhadra:1996mp}.  We then use the tuned value of $\kappa_b$ for the $B\to\pi\ell\nu$ form-factor production runs.  Table~\ref{tab:MILC_ens} shows the values of the clover coefficient and tuned $\kappa_b$ used in our calculation.

In order to take advantage of the improved action in the calculation of the $B\to\pi\ell\nu$ form factor, we must also improve the flavor-changing vector current to the same order in the heavy-quark expansion.  We remove errors of  $\CO(1/m_Q)$ in the vector current by rotating the heavy-quark field used in the matrix element calculation as
\begin{equation}
	 \psi_b \longrightarrow \Psi_b = \left( 1 + a\, d_1  \vec{\gamma} \cdot \vec{D}_\textrm{lat} \right) \psi_b ,
\label{eq:Psi_rot}
\end{equation}
where $\vec{D}_{\textrm{lat}}$ is the symmetric, nearest-neighbor, covariant difference operator.  We set $d_1$ to its tadpole-improved tree-level value of~\cite{ElKhadra:1996mp}
\begin{eqnarray} 
d_1=\frac{1}{u_0}\left(\frac{1}{2+m_0 a}-\frac{1}{2(1+m_0 a)}\right).
\end{eqnarray}
The values of the rotation parameter used in our calculation are given in Table~\ref{tab:MILC_ens}.

In order to convert dimensionful quantities determined in our lattice simulations into physical units, we need to know the value of the lattice spacing, $a$, which we find by computing a physical quantity that can be compared directly with experiment.  We first determine the relative lattice scale by calculating the ratio $r_1/a$ on each ensemble, where $r_1$ is related to the force between static quarks, $r_1^2F(r_1)=1.0$~\cite{Sommer:1993ce}.  These $r_1/a$ estimates are then smoothed by fitting to a smooth function of the gauge coupling and quark masses.  This scale-setting method has the advantage that the ratio $r_1/a$ can be determined precisely from a fit to the static quark potential~\cite{Bernard:2000gd,Aubin:2004wf}.   We convert all of our data from lattice spacing units into $r_1$ units before performing any chiral fits in order to account for slight differences in the value of the lattice spacing between ensembles.  In this work we use the value of $r_1^{\rm phys}=0.3108(15)(^{+26}_{-79})$ obtained by combining a recent lattice determination of $r_1 f_\pi$~\cite{Bernard:2007ps} with the PDG value of $f_\pi = 130.7 \pm 0.1 \pm 0.36$ MeV~\cite{Yao:2006px} to convert lattice results from $r_1$ units to physical units.

\subsection{Heavy-light meson correlation functions}
\label{sec:HLCorrs}

The $B\rightarrow\pi l\nu$ semileptonic form factors parameterize the hadronic matrix element of the $b \to u$ quark flavor-changing vector current $\CV^\mu\equiv i{\bar u}\gamma^\mu b$: 
\begin{equation}
\langle \pi | \CV^\mu | B \rangle  = f_+(q^2) \left( p^\mu_B + p^\mu_\pi - \frac{m_B^2 - m_\pi^2}{q^2}\,q^\mu \right) + f_0(q^2) \frac{m_B^2 - m_\pi^2}{q^2}\,q^\mu ,
\end{equation}
where $q^2$ is the momentum transferred to the outgoing lepton pair.   For calculations on the lattice and in HQET, it is more convenient to write the matrix element as~\cite{ElKhadra:2001rv}
\begin{equation}
\langle \pi | \CV^\mu | B \rangle  = \sqrt{2m_B} \left[ v^\mu f_\parallel(E_\pi) + p_\perp^\mu f_\perp(E_\pi) \right],
\end{equation}
where $v^\mu = p^\mu_B / m_B$ is the four-velocity of the $B$-meson, $p_\perp^\mu = p_\pi^\mu - (p_\pi \cdot v)v^\mu$ is the component of the pion momentum orthogonal to $v$, and $E_\pi = p_\pi \cdot v = (m_B^2 + m_\pi^2 - q^2)/(2 m_B)$ is the energy of the pion in the $B$-meson rest frame ($\vec{p}_B = \vec{0}$).  In this frame the form factors $f_\parallel(E_\pi )$ and $f_\perp(E_\pi )$ are directly proportional to the hadronic matrix elements:
\begin{eqnarray}
f_\parallel(E_\pi) & = & \frac{\langle \pi | \CV^0 | B \rangle}{\sqrt{2 m_B}} , \label{eq:fparallel}\\
f_\perp(E_\pi) & = & \frac{\langle \pi | \CV^i | B \rangle}{\sqrt{2 m_B}} \frac{1}{p_\pi^i}.\label{eq:fperp}
\end{eqnarray}
We therefore first calculate the hadronic matrix elements in Eqs. (\ref{eq:fparallel}) and (\ref{eq:fperp}) in the rest frame of the $B$-meson to obtain $f_\parallel(E_\pi )$ and $f_\perp(E_\pi )$, and then extract the standard form factors $f_0(q^2)$ and $f_+(q^2)$ using the following relations:
\begin{eqnarray}
	f_0 (q^2) & = & \frac{\sqrt{2 m_B}}{m_B^2 - m_\pi^2} \left[ (m_B - E_\pi) f_\parallel(E_\pi) + (E_\pi^2 - m_\pi^2) f_\perp(E_\pi) \right] , \\
	f_+ (q^2) & = & \frac{1}{\sqrt{2 m_B}} \left[ f_\parallel (E_\pi) + (m_B - E_\pi) f_\perp (E_\pi) \right].
\end{eqnarray}
These relations automatically satisfy the kinematic constraint $f_+(0) = f_0(0)$.

The 2-point and 3-point correlation functions needed to extract the lattice matrix element for $B\to\pi\ell\nu$ decay are
\begin{eqnarray}
	C_2^\pi(t; \vec{p}_\pi) & = & \sum_{\vec{x}} e^{i \vec{p}_\pi \cdot \vec{x}} \langle \CO_\pi (0,\vec{0})\, \CO^\dagger_\pi(t,\vec{x}) \rangle , \\
	C_2^B(t) & = & \sum_{\vec{x}} \langle \CO_B (0,\vec{0})\, \CO^\dagger_B(t,\vec{x}) \rangle , \\
	C_{3,\mu}^{B\to\pi}(t,T; \vec{p}_\pi) & = & \sum_{\vec{x},\vec{y}} e^{i \vec{p}_\pi \cdot \vec{y}} \langle \CO_\pi (0,\vec{0})\, V_\mu (t,\vec{y})\, \CO^\dagger_B (T,\vec{x}) \rangle ,
\end{eqnarray}
where $\CO_B$ and $\CO_\pi$ are interpolating operators for the $B$-meson and pion, respectively, and $V_\mu$ is the heavy-light vector current on the lattice.

In practice, to construct the heavy-light bilinears we must  combine a staggered light quark, which is a 1-component spinor, with a 4-component Wilson-type bottom quark;  we do so using the method established by Wingate \etal\ in Ref.~\cite{Wingate:2002fh}.  
For the $B$ meson we use a mixed-action interpolating operator:
\begin{equation}
        \mathcal{O}_{B,\Xi}(x) = \bar{\psi}_\alpha(x) \gamma^5_{\alpha\beta}
                \Omega_{\beta \Xi}(x) \chi(x),
        \label{eq:mixed-B-meson}
\end{equation}
where $\alpha,\beta$ are spin indices and $\Omega(x) \equiv \gamma_0^{x_0} \gamma_1^{x_1} \gamma_2^{x_2} \gamma_3^{x_3}$. The fields $\psi$ and $\chi$ are the 4-component clover quark field and 1-component staggered field, respectively.  Based on the transformation
properties of $\CO_{B,\Xi}(x)$ under shifts by one lattice spacing,  
$\Xi$ plays the role of a (fermionic) taste index~\cite{Wingate:2002fh,Kronfeld:2007ek}.  Once $\CO_{B,\Xi}(x)$ is summed over $2^4$ hypercubes in the correlation
functions that we compute, $\Xi$ also takes on the role of a taste degree of freedom, in the
sense of Refs.~\cite{Gliozzi:1982ib,KlubergStern:1983dg}.  Because the heavy quark field $\bar \psi_\alpha(x)$ is slowly varying over a hypercube, it does not affect the construction
of Refs.~\cite{Gliozzi:1982ib,KlubergStern:1983dg}.

 For the pion we use the local pseudoscalar interpolating operator,
\begin{equation}
        \mathcal{O}_\pi(x) = \varepsilon(x) \bar{\chi}(x)\chi(x),
        \label{eq:staggered-pion}
\end{equation}
where $\varepsilon(x) \equiv (-1)^{(x_1+x_2+x_3+x_4)}$.  We take the vector current to be
\begin{equation}
        V^\mu_\Xi(x) = \bar{\Psi}_\alpha(x) \gamma^\mu_{\alpha\beta}
                \Omega_{\beta \Xi}(x) \chi(x),
        \label{eq:mixed-vector-current}
\end{equation}
where $\Psi$ is the rotated heavy-quark field given in Eq.~(\ref{eq:Psi_rot}).  When forming $C_2^B(t)$ and $C_{3,\mu}^{B\to\pi}(t,T; \vec{p}_\pi)$, we sum over the taste index.  This yields the same correlation functions, with respect to taste, as in Ref.~\cite{Wingate:2002fh}.  Our principal difference with Ref.~\cite{Wingate:2002fh} is to use 4-component heavy quarks instead of 2-component non-relativistic quarks, and to derive the correlators in the staggered formalism, without the introduction of naive fermions.

We work in the rest frame of the $B$-meson, so only the pions carry momentum.  We compute both the 2-point function $C_2^\pi(t; \vec{p}_\pi)$ and the 3-point function $C_{3,\mu}^{B\to\pi}(t,T; \vec{p}_\pi)$ at discrete values of the momenta $\vec{p}_\pi = 2\pi(0,0,0)/L, 2\pi(1,0,0)/L, 2\pi(1,1,0)/L,2\pi(1,1,1)/L$, and $2\pi(2,0,0)/L$ allowed by the finite spatial lattice volume.  We use only data through momentum $\vec{p}_\pi = 2\pi(1,1,1)/L$, however, because the statistical errors in the correlators increase significantly with momentum.

We use a local source for the pions throughout the calculation, while we smear the $B$-meson wavefunction in both the 2-point function $C_2^B(t) $ and the 3-point function $C_{3,\mu}^{B\to\pi}(t,T; \vec{p}_\pi)$:
\begin{equation}
	\widetilde{\mathcal{O}}_{B,\Xi}(t, \vec{x}) = \sum_{\vec{y}} S(\vec{y}) \bar{\psi}_\alpha(t, \vec{x} + \vec{y}) \gamma^5_{\alpha\beta} \Omega_{\beta \Xi}(t, \vec{x}) \chi(t, \vec{x}),
\end{equation}
where $S(\vec{y})$ is the spatial smearing function.  This reduces contamination from heavier excited states and allows a better determination of the desired ground state amplitude.   In our study of choices for how to smear the $B$-meson, we found that a wall source, $S(\vec{y})=1$, worked extremely well for suppressing excited state contamination, but at the cost of large statistical errors in the 2-point and 3-point correlation functions.  In contrast, use of a 1S wavefunction, $S(\vec{y}) = \text{exp}(- \mu |\vec{y}|)$, optimized to have good overlap with the charmonium ground state led to smaller statistical errors at the cost of undesirably large excited state contributions to the 3-point function that would make it difficult to extract the ground state amplitude.  In order to achieve a balance between small statistical errors and minimal excited state contamination, we tune the coefficient of the exponential in the 1S wavefunction to the smallest value (\ie the widest smearing) for which the $B$-meson 2-point effective mass is still well-behaved; we find a value of $a \mu=0.20$ for the coarse ensembles.   We note that our determination of the optimal $B$-meson smearing function is consistent with the theoretical expectation that the $B$-meson wavefunction should be wider than the charmonium wavefunction.

For the calculation of the 3-point function, we fix the location of the pion source at $t_i=0$ and the location of the $B$-meson sink at $t_f=T$, and vary the position of the operator over all times $t$ in between.  If the source-sink separation is too small then the entire time range $0 < t < T$ is contaminated by excited states, but if the source-sink separation is too large then the correlation function becomes extremely noisy. In practice, we set the sink time to $T=16$ on the coarse lattices;  we have checked, however, that our results using this choice are consistent with those determined from using $T=12$ and $T=20$.  On the fine lattices we scale the source sink separation by the approximate ratio of the lattice spacings, $a_\textrm{fine}/a_\textrm{coarse}$, and use $T=24$.

In order to minimize the statistical errors given the available number of configurations in each ensemble, we compute the necessary 2-point and 3-point correlations not only with a source time of $t_i = 0$, but also with source times of $t_i = n_t/4, n_t/2$, and $3n_t/4$ ($n_t$ is the temporal extent of the lattice) and the sink time $T$ shifted accordingly.  We then average the results from the four source times; this effectively increases our statistics by a factor of four.

\subsection{Heavy-light current renormalization}
\label{sec:HLCurrents}

In order to recover the desired continuum matrix element, the lattice amplitude must be multiplied by the appropriate renormalization factor $Z^{bl}_{V_\mu}$:
\begin{equation}
	\langle \pi | \CV_\mu | B \rangle = Z^{bl}_{V_\mu} \times \langle \pi | V_\mu | B \rangle ,
\label{eq:V_renorm}
\end{equation}
where $V_\mu$ and $\CV_\mu$ are the lattice and continuum $b\to u$ vector currents, respectively. This removes the dominant discretization errors from the lattice current operator.  In terms of the form factors, Eq.~(\ref{eq:V_renorm}) can be rewritten as
\begin{eqnarray}
	f_\parallel & = & Z^{bl}_{V_0} \times  f^\textrm{lat}_\parallel \\
	f_\perp & = & Z^{bl}_{V_i} \times f^\textrm{lat}_\perp ,
\end{eqnarray}
where explicit expressions relating $f^\textrm{lat}_\parallel$ and $f^\textrm{lat}_\perp$ to correlation functions are given in Eqs.~(\ref{eq:fpar_R})~and~(\ref{eq:fperp_R}).

In this work, we calculate $Z^{bl}_{V_\mu}$ via the mostly nonperturbative method used in the earlier quenched Fermilab calculation~\cite{ElKhadra:2001rv}.  We first rewrite $Z^{bl}_{V_\mu}$ as
\begin{eqnarray}
	Z^{bl}_{V_\mu} = \rho_{V_\mu}^{hl}  \sqrt{Z_V^{bb} Z_V^{ll} } .
\label{eq:renorm}
\end{eqnarray}
The flavor-conserving renormalization factors $Z_V^{bb}$ and $Z_V^{ll}$ account for most of the value of $Z^{bl}_{V}$~\cite{Harada:2001fi}.  They can be determined from standard heavy-light meson charge normalization conditions:
\begin{eqnarray}
	Z_V^{ll}  \times  \langle D |  V^{{ll},0} | D \rangle &=& 1, \label{eq:ZVll} \\
	Z_V^{bb}  \times  \langle B | V^{{bb},0} | B \rangle &=& 1, \label{eq:ZVhh}
\end{eqnarray}
where the light-light and heavy-heavy lattice vector currents are given by
\begin{eqnarray}
	V^{{ll},\mu}_{\Xi\Xi^\prime}(x) &=& \chi^\dagger(x) \Omega(x)^\dagger_{\Xi\alpha} \gamma^\mu_{\alpha\beta} \Omega(x)_{\beta\Xi^\prime} \chi(x) , \\
	V^{{bb},\mu}(x) &=& {\bar{\Psi}_b}_\alpha(x) \gamma^\mu_{\alpha\beta} {\Psi_b}_\beta(x),
\end{eqnarray}
respectively.  In order to reduce the statistical errors in $Z_V^{ll}$, we compute the lattice matrix element $\langle D |  V^{{ll},0} | D \rangle$ using a clover charm quark as the spectator in the 3-point correlation function.  We eliminate contamination from staggered oscillating states in the determination of $Z_V^{bb}$ by using a clover strange quark for the spectator in the 3-point correlation function $ \langle B | V^{{bb},0} | B \rangle$.  Once $Z_V^{ll}$ and $Z_V^{bb}$ have been determined nonperturbatively, the remaining correction factor in Eq.~(\ref{eq:renorm}), $\rho_{V_\mu}^{hl}$, is expected to be close to unity because most of the radiative corrections, including contributions from tadpole graphs, cancel in the ratio~\cite{Harada:2001fi}.  We therefore estimate $\rho_{V_\mu}^{hl}$ from 1-loop lattice perturbation theory~\cite{Lepage:1992xa}.

The matching factor $\rho_{V_\mu}^{hl}$ has been calculated by a subset of the present authors, and a separate publication describing the details is in preparation~\cite{ElKhadra_PT}.  The corrections to $\rho_{V_\mu}^{hl}$ can be expressed as a perturbative series expansion in powers of the strong coupling:
\begin{equation}
	\rho_{V_\mu}^{hl}  = 1 + 4\pi \alpha_V(q^*) \rho_{V_\mu}^{hl[1]} + \CO(\alpha_V^2) ,
\end{equation}
where $\alpha_V(q^*)$ is the renormalized coupling constant in the $V$-scheme and is determined from the static quark potential with the same procedure as is used in Ref.~\cite{Mason:2005zx}.  The scale $q^*$, which should be the size of a typical gluon loop momentum, is computed via an extension of the methods outlined by Brodsky, Lepage, and Mackenzie~\cite{Lepage:1992xa,Brodsky:1982gc} and Hornbostel, Lepage, and Morningstar~\cite{Hornbostel:2002af}.  The value of $q^*$ ranges from 2.0--4.5 GeV for the parameters used in our simulations.   The 1-loop coefficient, $\rho_{V_\mu}^{hl[1]}$, and higher moments are calculated using automated perturbation theory and numerical integration as described in Refs.~\cite{Luscher:1985wf,ElKhadra:2007qe}.  We find that the perturbative corrections to matrix elements of the temporal vector current, $V_0$, are less than a percent, while the corrections to matrix elements of the spatial vector current, $V_i$, are 3--4\%.

\section{Analysis}
\label{sec:Analysis}

In this section, we describe the three-step analysis procedure used to calculate the $B\to\pi\ell\nu$ semileptonic form factor, $f_+(q^2)$.  In the first subsection, Sec.~\ref{sec:2point}, we describe how we fit the pion and $B$-meson 2-point correlation functions in order to determine the pion energies and $B$-meson mass.  We use both of these quantities in the later determination of the lattice form factors $f_\parallel (E_\pi)$ and $f_\perp (E_\pi)$.  Next, in Sec.~\ref{sec:3point}, we construct a useful ratio of the  3-point correlation function $\langle \pi | V | B \rangle$ to the 2-point functions.  We then fit this ratio to a simple plateau ansatz to extract the desired form factors.  Finally, in Sec.~\ref{sec:ChPT}, we extrapolate the form factors calculated at unphysically heavy quark masses and finite lattice spacing to the physical light quark masses and zero lattice spacing using next-to-leading order (NLO) HMS$\chi$PT expressions extended with next-to-next-to-leading order (NNLO) analytic terms.  (We perform a simultaneous extrapolation in $m_q$ and $a$ and interpolation in $E_\pi$.)  We then take the appropriate linear combination of $f_\parallel (E_\pi)$ and $f_\perp (E_\pi)$ to determine the desired form factor,  $f_+(q^2)$, with statistical errors.

\subsection{Two-point correlator fits}
\label{sec:2point}

The pion and $B$-meson 2-point correlators obey the following functional forms:
\begin{eqnarray}
	C_{2}^{\pi} (t; \vec{p}_\pi) & = & \sum_{m} (-1)^{mt} \, | \langle 0 | \CO_\pi | \pi^{(m)} \rangle |^2 \, \frac{e^{-E_\pi^{(m)}t}}{2 E_\pi^{(m)}} , \\
	C_{2}^{B} (t) & = & \sum_{m} (-1)^{mt} \, | \langle 0 | \CO_B | B^{(m)} \rangle |^2 \, \frac{e^{-m_B^{(m)}t}}{2 m_B^{(m)}} .
\end{eqnarray}
In the above expressions, terms with odd $m$ contain the prefactor $(-1)^t$.  This leads to visible oscillations in time in the meson propagators; such behavior arises with staggered quarks because the parity operator is a composition of spatial inversion and translation through one timeslice~\cite{Golterman:1984cy,Kilcup:1986dg}.  The contributions of the opposite-parity oscillating states are found to be significant throughout the entire time range and must therefore be included in fits to extract the desired ground state energy.  

Because the statistical errors in the pion energies and $B$-meson mass contribute very little to the total statistical error in the $B\to\pi\ell\nu$ form factor, we use a simple procedure to fit the 2-point functions.  Although this does not optimize the determinations of $E_\pi$ and $m_B$, it is sufficient for the purpose of this analysis.  We first select a fit range, $t_\textrm{min} $--$t_\textrm{max}$, that allows a good correlated, unconstrained fit including only contributions from the ground state and its opposite-parity partner. We then reduce $t_\textrm{min}$ by one timeslice and redo the fit.  If the correlated confidence level is too low ($\ltapprox 10\%$), we increase the number of states and try the fit again with the same time range.  Otherwise, if the fit is good, we reduce $t_\textrm{min}$ by one more timeslice and repeat the fit.  We repeat this procedure until we can no longer get a good fit without using a large number (greater than 4) of states.  We note that, by including only as many states as the data can determine, we minimize the possibility of spurious solutions in which the fitter exchanges the ground state with one of the same-parity excited states.  We have, however, checked that this method yields the same results within statistical errors as a constrained fit that includes up to three or four pairs of states.  

Figure~\ref{fig:E_vs_tmin} shows examples of both $m_\pi$ vs. $t_\textrm{min}$ (left plot) and $m_B$ vs. $t_\textrm{min}$ (right plot) on the $am_l/am_s = 0.02/0.05$ coarse ensemble, which has the largest light quark mass of the coarse ensembles.  The masses are stable as $t_\textrm{min}$ is reduced, and the statistical errors in $m_B$ become smaller as additional timeslices are added to the fit.  The statistical errors are determined by performing a separate fit to 500 bootstrap ensembles;  each fit uses the full single elimination jackknife correlation matrix which is remade before every fit.  The size of the statistical errors does not change when the number of bootstrap ensembles is increased by factors of two or four.  We select the time range to use in the $B\to\pi\ell\nu$ analysis based on several criteria:  a good correlated confidence level, relatively symmetric upper and lower bootstrap error bars, no $5$-$\sigma$ or greater outliers in the bootstrap distribution, and no sign of excited state contamination.  The red (filled) data points in Fig.~\ref{fig:E_vs_tmin} mark the chosen fit ranges for the ensemble in the example plots.  Figures~\ref{fig:pi_2pt_fits} and~\ref{fig:B_2pt_fits} show the corresponding pion and $B$-meson correlator fits, respectively, which go through the data points (shown with jackknife errors) quite well.

\begin{CFigure}
	\begin{tabular}{cc}
	\hspace{-0.00 \textwidth}
	\rotatebox{0}{\includegraphics[width=0.47\textwidth]{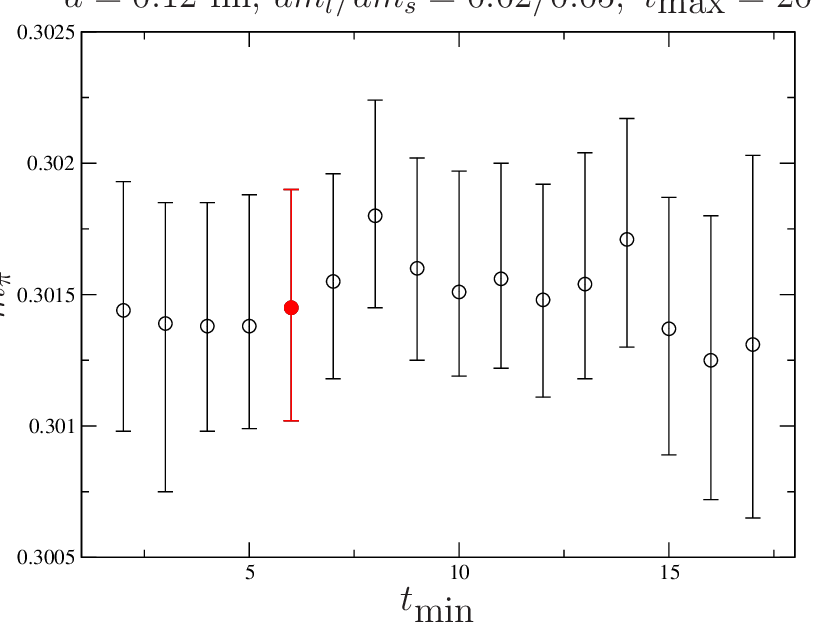}} 
	& \hspace{0.03 \textwidth}
	\rotatebox{0}{\includegraphics[width=0.47\textwidth]{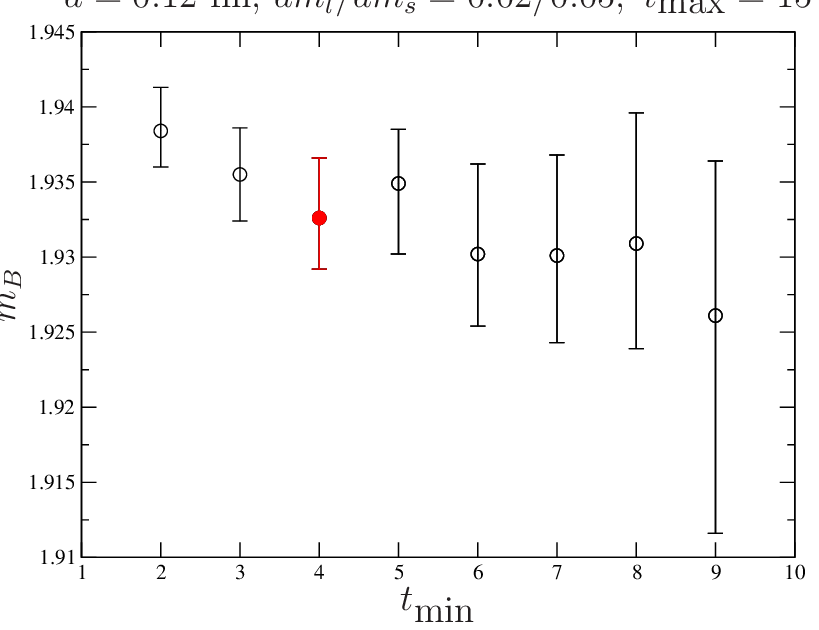}}
	\end{tabular}
	\caption{Pion mass (left plot) and $B$-meson mass (right plot) versus minimum timeslice in 2-point correlator fit.  The red (filled) data points show the fit ranges selected for use in the $B\to\pi\ell\nu$ form factor analysis.}
	\label{fig:E_vs_tmin}
\end{CFigure}

\begin{CFigure}
	\vspace{0.02\textwidth}
	\begin{tabular}{cc}
	\hspace{0.00 \textwidth}
	\rotatebox{0}{\includegraphics[width=0.47\textwidth]{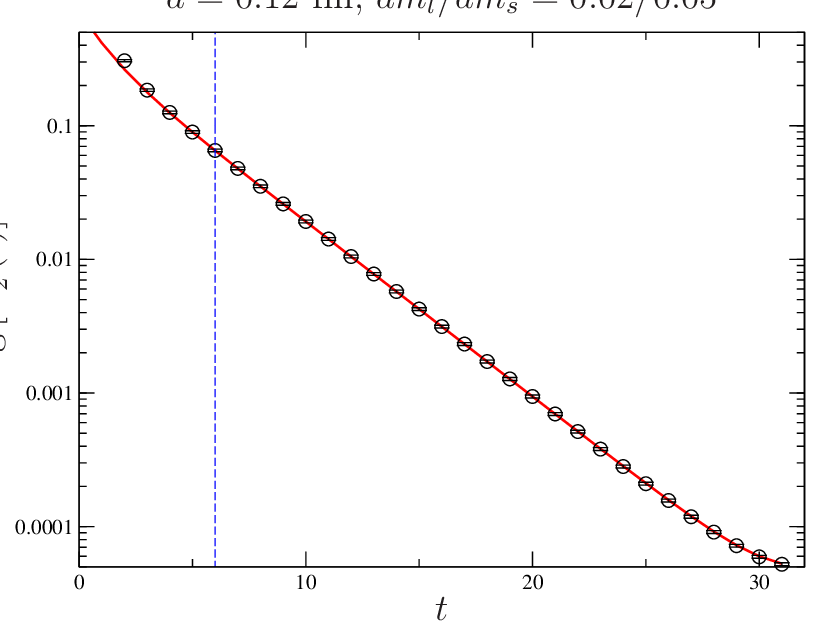}}
	& \hspace{0.03 \textwidth}
	\rotatebox{0}{\includegraphics[width=0.47\textwidth]{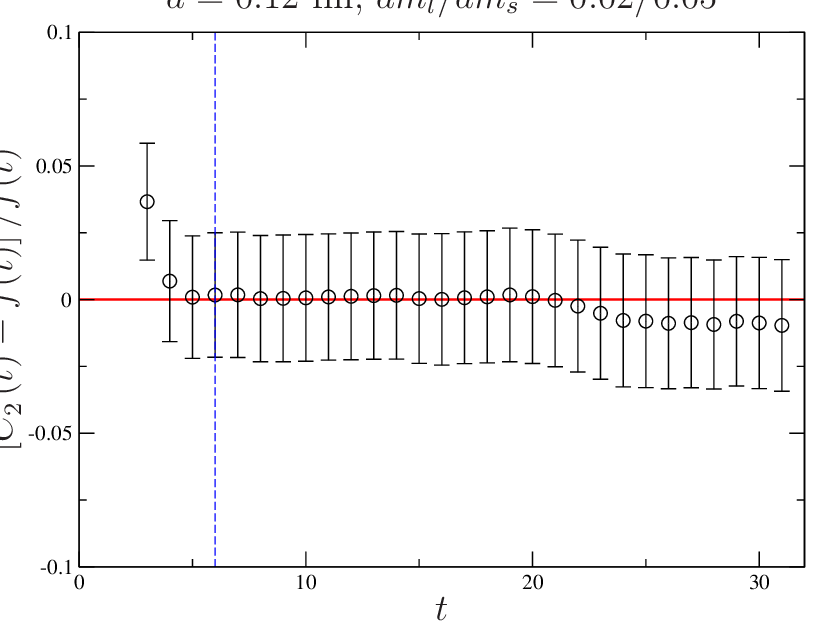}}
	\end{tabular}
	\caption{Pion correlator fit corresponding to the red data point in the left-hand graph of Fig.~\ref{fig:E_vs_tmin}.  The left plot shows the fit (red line) to the zero-momentum pion propagator on a log scale, while the right plot shows the deviation of the fit from the data point for each timeslice. On both plots the dashed vertical line indicates $t_\text{min}$.  Single elimination jackknife statistical errors are shown.}
	\label{fig:pi_2pt_fits}
\end{CFigure}
\begin{CFigure}
	\begin{tabular}{cc}
	\hspace{0.0 \textwidth}
	\rotatebox{0}{\includegraphics[width=0.47\textwidth]{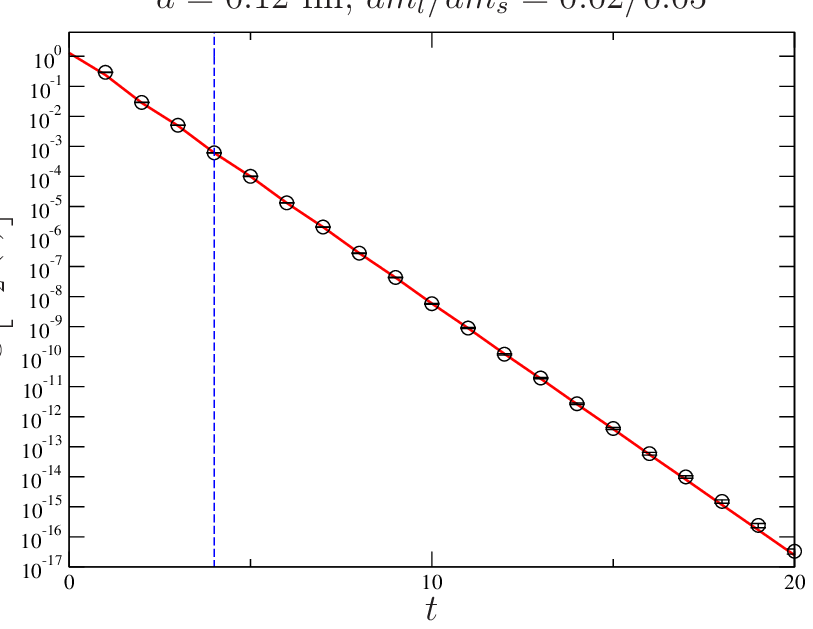}} 
	& \hspace{0.03 \textwidth}
	\rotatebox{0}{\includegraphics[width=0.47\textwidth]{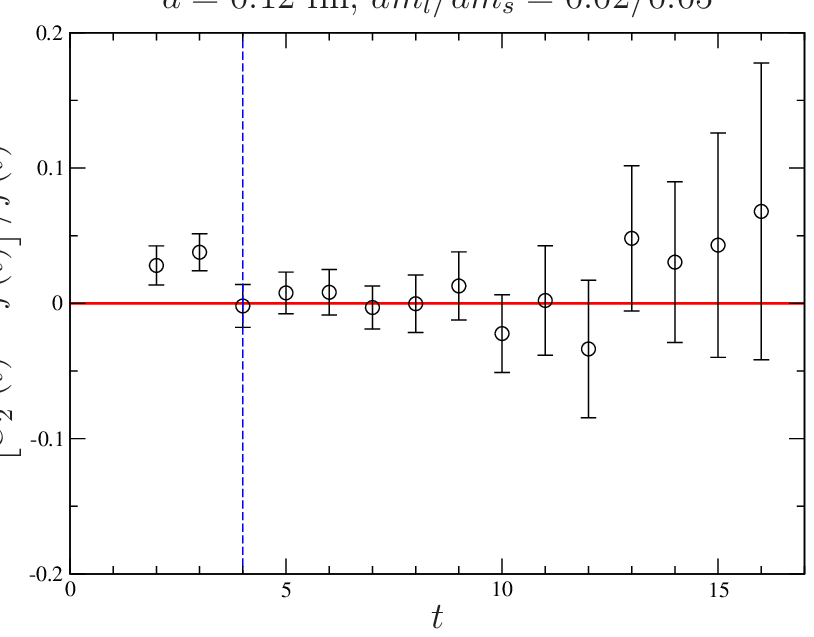}}
	\end{tabular}
	\caption{$B$-meson correlator fit corresponding to the red data point in the right-hand graph of Fig.~\ref{fig:E_vs_tmin}.  The left plot shows the fit (red line) to the $B$-meson propagator on a log scale, while the right plot shows the deviation of the fit from the data point for each timeslice. On both plots the dashed vertical line indicates $t_\text{min}$.  Single elimination jackknife statistical errors are shown.}	
\label{fig:B_2pt_fits}
\end{CFigure}

The gauge configurations have been recorded every six trajectories, and the remaining autocorrelations between consecutive configurations cannot be neglected.  We address this by averaging a block of successive configurations together before calculating the correlation matrix and performing the fit.  We determine the optimal block size by increasing the number of configurations in a block until the single elimination jackknife statistical error in the correlator data remains constant within errors.  This is shown for a representative timeslice of the pion propagator on a coarse ensemble in Fig.~\ref{fig:blocksize}.  We find that it is necessary to use a block size of 5 on the coarse ensembles and 8 on the fine ensembles, and we use these values for the rest of the form factor analysis.  We note that the size of the statistical errors that arises from blocking by 5 on the coarse ensemble is consistent with that estimated based on a calculation of the integrated autocorrelation time.

\begin{CFigure}
	\vspace{0.02\textwidth}
	\rotatebox{0}{\includegraphics[width=0.47\textwidth]{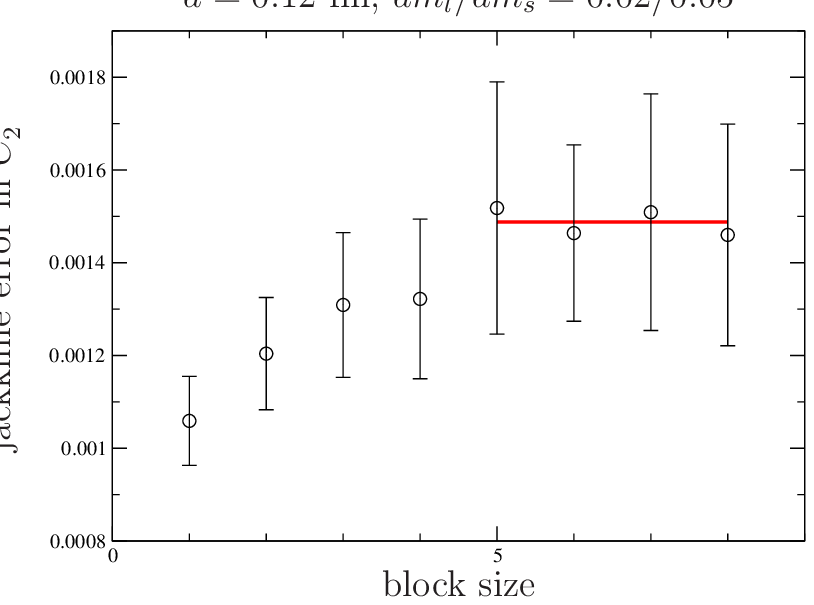}}
	\caption{Single-elimination jackknife error versus block size in the zero-momentum pion propagator at $t=6$.  The statistical errors in the errors are calculated with an additional single elimination jackknife loop.  The red line is an average of the errors for block sizes 5--8 and is only to make it easier to see that the statistical error plateaus after a block size of 5;  it is not used in the form factor analysis.}
	\label{fig:blocksize}
\end{CFigure}

The pion energy $E_\pi$ that is extracted from fitting the 2-point function, $C_{2}^{\pi} (t; \vec{p}_\pi)$, should satisfy the dispersion relation $E_\pi^2  =  |\vec{p}_\pi |^2 + m_\pi^2 $ in the continuum limit due to the restoration of rotational symmetry.  Similarly, the pion amplitude, $Z_\pi \equiv | \langle 0 | \CO_\pi | \pi \rangle |$, should be independent of $\vec{p}_\pi$ as $a\to 0$.  As shown in Fig.~\ref{fig:pi_disp_check}, our results are consistent with these continuum relations within statistical errors.\footnote{As this analysis was being completed we generated data with four times the statistics on the $am_l/am_s = 0.02/0.05$  coarse ensemble.  In order to make the comparison to the continuum expectation clearer, we use the higher statistics data in Fig.~\ref{fig:pi_disp_check}.}  We therefore replace the pion energy $E_\pi$ by $\sqrt{|\vec{p}_\pi |^2 + m_\pi^2}$ when calculating the lattice form factors $f_{\parallel}(E_\pi)$ and $f_{\perp}(E_\pi)$ in order to reduce the total statistical uncertainty.  The pion amplitude drops out of the form factor calculation, however, because we take suitable ratios of 3-point to 2-point correlators.

\begin{CFigure}
	\begin{tabular}{cc}
	\hspace{-0.00 \textwidth}
	\rotatebox{0}{\includegraphics[width=0.47\textwidth]{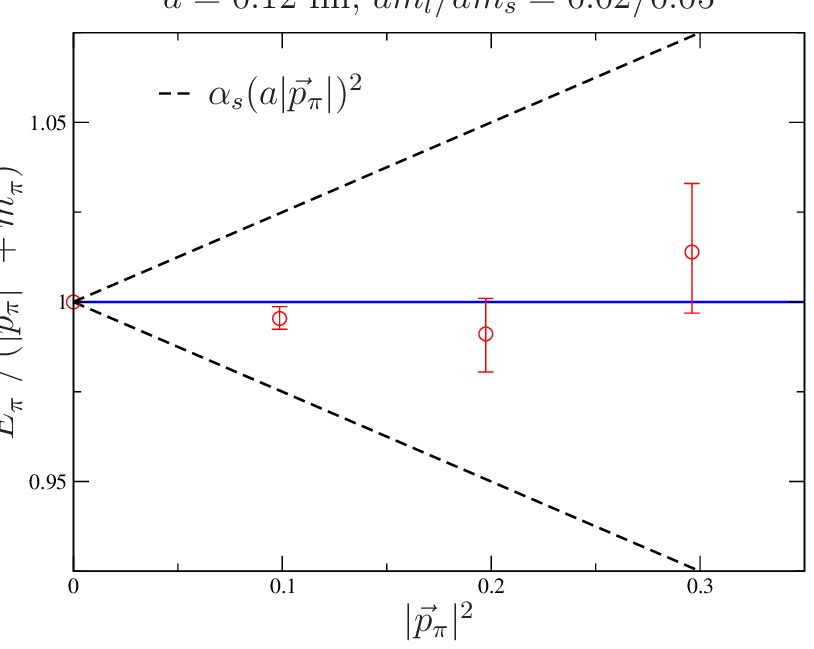}} 
	&  \hspace{0.03 \textwidth}
	\rotatebox{0}{\includegraphics[width=0.47\textwidth]{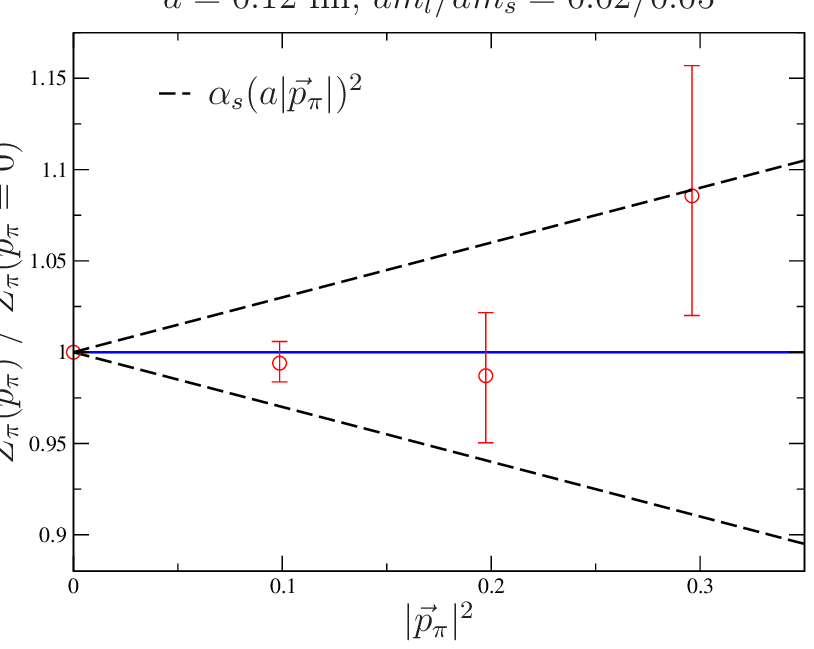}}
	\end{tabular}
	\caption{Comparison of pion energy $E_\pi$ (left plot) and amplitude $Z_\pi$ (right plot) with the prediction of the continuum dispersion relation.  We also show a power-counting estimate for the size of momentum-dependent discretization errors, which are of $\CO(\alpha_s (a |\vec{p}_\pi |)^2)$, as dashed black lines.}
\label{fig:pi_disp_check}
\end{CFigure}

\subsection{Three-point correlator fits}
\label{sec:3point}

The $B\to\pi$ 3-point correlator obeys the following functional form:
\begin{eqnarray}
	C_{3,\mu}^{B\to \pi} (t,T) & = & \sum_{m,n} (-1)^{mt} (-1)^{n(T-t)} A_{\mu}^{mn} e^{-E_\pi^{(m)}t} e^{-m_B^{(n)}(T-t)} ,
\end{eqnarray}
where
\begin{eqnarray}
	A_{\mu}^{mn} & \equiv &  \frac{\langle 0 | \CO_\pi | \pi^{(m)} \rangle}{2 E_\pi^{(m)}} \langle \pi^{(m)} | V_\mu | B^{(n)} \rangle  \frac{\langle B^{(n)} | \CO_B | 0 \rangle}{2 m_B^{(n)}} .
\label{eq:3pt}\end{eqnarray}
Writing out the first four terms of $C_{3,\mu}^{B\to \pi} (t,T)$ makes the behavior of the 3-point correlator as a function of both $t$ and $T$ more transparent:
\begin{eqnarray}
	C_{3,\mu}^{B\to \pi} (t,T) & = & A_{\mu}^{00} e^{-E_\pi^{(0)}t} e^{-m_B^{(0)}(T-t)} +  (-1)^{(T-t)} A_{\mu}^{01} e^{-E_\pi^{(0)}t} e^{-m_B^{(1)}(T-t)} \nonumber\\ 
	& + &  (-1)^{t} A_{\mu}^{10} e^{-E_\pi^{(1)}t} e^{-m_B^{(0)}(T-t)}  + (-1)^{T} A_{\mu}^{11} e^{-E_\pi^{(1)}t} e^{-m_B^{(1)}(T-t)} + \ldots 
\end{eqnarray}
As in the case of the pion and $B$-meson propagators, the leading contributions from the opposite-parity excited states (the $A_{\mu}^{10}$ and $ A_{\mu}^{01}$ terms) change sign when $t \to t+1$; these produce visible oscillations in the correlation function along the time direction.   The subleading contribution from the opposite-parity excited states (the $A_{\mu}^{11}$ term), however, only changes sign when the source-sink separation is varied, \emph{e.g.}, $T \to T+1$; this contribution is not as clearly visible in the data as those that oscillate with the time slice $t$.

The lattice form factors are related to the ground-state amplitude of the 3-point function $C_{3,\mu}^{B\to \pi} (t,T)$ as follows:
\begin{eqnarray}
	f_\parallel^\textrm{lat} & = & A^{00}_0 \left( \frac{2 E_\pi \sqrt{2 m_B}}{Z_\pi Z_B} \right) \\
	f_\perp^\textrm{lat} & = & A^{00}_i \left(\frac{2 E_\pi \sqrt{2 m_B}}{Z_\pi Z_B}\right) \frac{1}{p^i_\pi} ,
\end{eqnarray}
where, as before, $Z_\pi \equiv | \langle 0 | \CO_\pi | \pi \rangle |$ and $Z_B \equiv | \langle 0 | \CO_B | B \rangle |$.  The pion and $B$-meson energies and amplitudes are known from the 2-point fits described in the previous subsection.  Thus, the goal is to determine the 3-point amplitude $A^{00}_\mu$ for $\mu$ along both the spatial and temporal directions.

In principle, the easiest way to determine the coefficient $A^{00}_\mu$ is to divide the 3-point function $C_{3,\mu}^{B\to \pi} (t,T)$ by the appropriate 2-point functions and fit to a constant (plateau) ansatz in a region of time slices $0 \ll t \ll T$ that are sufficiently far from both the pion and $B$-meson sources, such that excited state contamination can be neglected.  In practice, however, oscillating excited-state contributions are significant throughout the interval between the pion and $B$-meson, so our raw correlator data cannot be fit to such a simple function.  Therefore we construct an average correlator in which the oscillations are reduced before performing any fits.  This method for determining the form factors requires knowledge of $E_\pi$ and $m_B$;  we use the values determined in the 2-point fits described in the previous subsection and propagate the bootstrap uncertainties in order to properly account for correlations.

The final ratio of correlators used to determine $A^{00}_\mu$ entails several pieces.  To begin consider the carefully constructed average of the value of the $B$-meson propagator at time slice $t$ with that at $t+1$: 
\begin{eqnarray}
	C_{2}^{B} (t) \longrightarrow C_{2}^{'B} (t) &=& \frac{e^{-m_B^{(0)}t}}{2} \left[  \frac{C_{2}^{B} (t)}{e^{-m_B^{(0)}t}} + \frac{C_{2}^{B} (t+1)}{e^{-m_B^{(0)}(t+1)}}  \right] \nonumber\\
	&=& \frac{Z_B^2}{2 m^{(0)}_B} e^{-m_B^{(0)}t} +   (-1)^t \frac{Z_B'^2}{2 m^{(1)}_B} e^{-m_B^{(1)}t} \left( \frac{1 - e^{-\Delta m_B}}{2} \right) + \ldots ,
\label{eq:MO_ave}
\end{eqnarray}
where  $\Delta m_B \equiv m^{(1)}_B - m^{(0)}_B$.  By removing the leading exponential behavior from the correlator \emph{before} taking the average we suppress the leading oscillating contribution by a factor of the mass-splitting $\Delta m_B / 2$ while leaving the desired ground state amplitude unaffected.    Note also that, while this procedure affects the size of the excited state amplitudes, it does not alter the functional form of the correlator, nor does it alter the energies in the exponentials.  Therefore the average in Eq.~(\ref{eq:MO_ave}) is equivalent to using a smeared source that has a smaller coupling to the opposite-parity excited states.  This averaging procedure can be iterated in order to make the oscillating terms arbitrarily small.  Empirically, we find that two iterations are sufficient for all of our numerical data:
\begin{eqnarray}
	\bar{C}_{2}^{B} (t) &\equiv& \frac{e^{-m_B^{(0)}t}}{4} \left[ \frac{C_{2}^{B} (t)}{e^{-m_B^{(0)}t}} + \frac{2 C_{2}^{B} (t+1)}{e^{-m_B^{(0)}(t+1)}} + \frac{C_{2}^{B} (t+2)}{e^{-m_B^{(0)}(t+2)}}  \right] \nonumber\\
	&\approx& \frac{Z_B^2}{2 m^{(0)}_B} e^{-m_B^{(0)}t} +   (-1)^t \frac{Z_B'^2}{2 m^{(1)}_B} e^{-m_B^{(1)}t} \left( \frac{\Delta m_B^2}{4} \right)  + \CO(\Delta m_B^3) .
\label{eq:MO_iterave}
\end{eqnarray}
At our various light quark masses and lattice spacings  the mass-splittings lie in the range $0.1 \ltapprox \Delta m_B \ltapprox 0.3$ in lattice units;  thus use of the iterated average in Eq.~(\ref{eq:MO_iterave}) reduces the leading oscillating state amplitude by a factor of $\sim$50--400 such that it can be safely neglected.

In the case of the $B\to\pi$ 3-point correlation function, we wish to reduce both the oscillating contributions and the less visible non-oscillating contributions that arise from the cross-term between the lowest-lying pion and $B$-meson opposite-parity states.   If these contributions
are reduced sufficiently, we can safely neglect all of them when extracting the ground-state amplitude $A^{00}_\mu$.  We therefore construct a slightly more sophisticated average which combines the correlator both at consecutive time slices ($t$ and $t+1$) and at consecutive source-sink separations ($T$ and $T+1$):
\begin{eqnarray}
	\bar{C}_{3,\mu}^{B\to \pi} (t,T) & = &  \frac{e^{-E_\pi^{(0)}t} \, e^{-m_B^{(0)}(T-t)}}{8} \nonumber\\
		& \times & \bigg[ \frac{C_{3,\mu}^{B\to\pi} (t,T)}{e^{-E_\pi^{(0)}t} e^{-m_B^{(0)}(T-t)}} + \frac{C_{3,\mu}^{B\to\pi} (t,T+1)}{e^{-E_\pi^{(0)}(t)} e^{-m_B^{(0)}(T+1-t)}}  +  \frac{2 \, C_{3,\mu}^{B\to\pi} (t+1,T)}{e^{-E_\pi^{(0)}(t+1)} e^{-m_B^{(0)}(T-t-1)}} \nonumber\\
		& + & \frac{2 \, C_{3,\mu}^{B\to\pi} (t+1,T+1)}{e^{-E_\pi^{(0)}(t+1)} e^{-m_B^{(0)}(T-t)}} +  \frac{C_{3,\mu}^{B\to\pi} (t+2,T)}{e^{-E_\pi^{(0)}(t+2)} e^{-m_B^{(0)}(T-t-2)}} +  \frac{C_{3,\mu}^{B\to\pi} (t+2,T+1)}{e^{-E_\pi^{(0)}(t+2)} e^{-m_B^{(0)}(T-t-1)}} \bigg] \nonumber\\
		& \approx &  A_{\mu}^{00}e^{-E_\pi^{(0)}t} \, e^{-m_B^{(0)}(T-t)} + (-1)^{T} A_{\mu}^{11} e^{- E_\pi^{(1)}t} e^{-m_B^{(1)}(T-t)} \left(\frac{\Delta m_B}{2}  \right) \nonumber\\ & + & \CO(\Delta E_\pi^2,\, \Delta E_\pi \Delta m_B,\, \Delta m_B^2) .
\end{eqnarray}
This average reduces the unwanted parity states' contamination significantly.  It eliminates both the leading $\CO(1)$ and subleading $\CO(\Delta E_\pi)$ contributions to the oscillating $A^{10}$ term, the two lowest-order $\CO(1, \Delta m_B)$ contributions to the oscillating $A^{01}$ term, and the $\CO(1, \Delta E_\pi)$ contributions to the non-oscillating $A^{11}$ term.  The size of the remaining $A^{11}$ term is a factor of $\sim$7--20 times smaller than in the unsmeared 3-point correlator.

We can now safely ignore contamination from opposite-parity states and determine the lattice form factors in a simple manner.  We construct the following ratio of the smeared correlators:
\begin{eqnarray}
	\bar{R}_{3,\mu}^{B\to \pi} (t,T) & \equiv & \frac{\bar{C}_{3,\mu}^{B\to \pi} (t,T)}{\sqrt{\bar{C}_{2}^{\pi} (t)\bar{C}_{2}^{B} (T-t)}} \sqrt{\frac{2 E_\pi}{e^{-E_\pi^{(0)}t} \, e^{-m_B^{(0)}(T-t)}}} .
	\end{eqnarray}
The lattice form factors are then:
\begin{eqnarray}
\label{eq:fpar_R}
	f_\parallel^\textrm{lat} & = &\bar{R}_{3,0}^{B\to \pi} (t,T) \\
	f_\perp^\textrm{lat} & = & \frac{1}{p^i_\pi} \, \bar{R}_{3,i}^{B\to \pi} (t,T)  .
\label{eq:fperp_R}
\end{eqnarray}

We fit $f_\parallel^\textrm{lat}$ and $f_\perp^\textrm{lat}$ as defined in Eqs.~(\ref{eq:fpar_R})--(\ref{eq:fperp_R}) to a plateau in the region $0 \ll t \ll T$ where ordinary excited state contributions can be neglected.  Figure~\ref{fig:R_plateaus} shows the determinations of $f_\parallel^\textrm{lat}$ (left plot) and $f_\perp^\textrm{lat}$ (right plot) for all of the momenta that we use in the chiral extrapolation on the coarse ensemble with $am_l / am_s = 0.02/0.05$.  In practice, we fit a range of four time slices, choosing the interval that results in the best correlated confidence level.  We have cross-checked the determination of the form factors via Eqs.~(\ref{eq:fpar_R})--(\ref{eq:fperp_R}) against determinations of the form factor that explicitly include excited state dependence in the fit ansatz and find that the results agree within errors.  Our preferred method, however, yields the smaller statistical uncertainty in the form factors.
\begin{CFigure}
	\begin{tabular}{cc}
	\hspace{0.0 \textwidth}
	\rotatebox{0}{\includegraphics[width=0.47\textwidth]{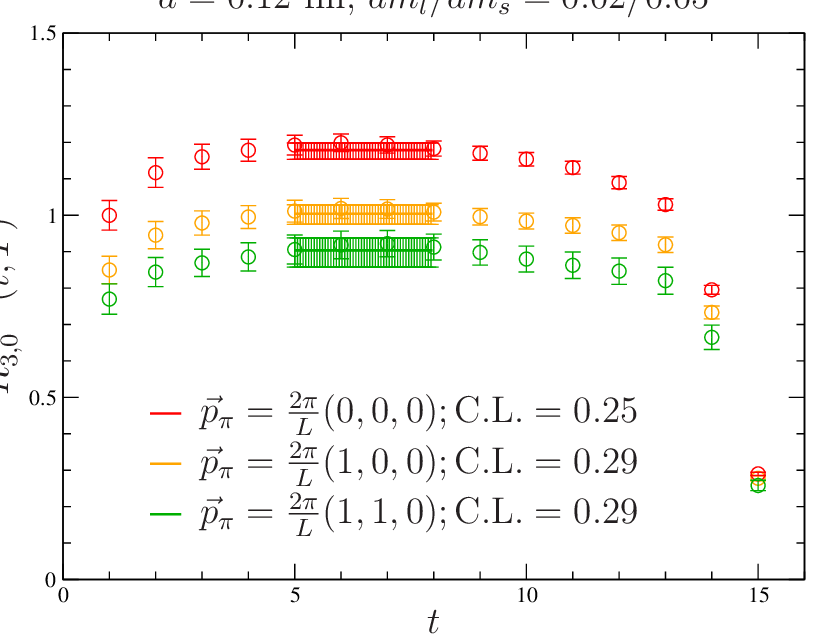}} 	
	& \hspace{0.035 \textwidth}
	\rotatebox{0}{\includegraphics[width=0.47\textwidth]{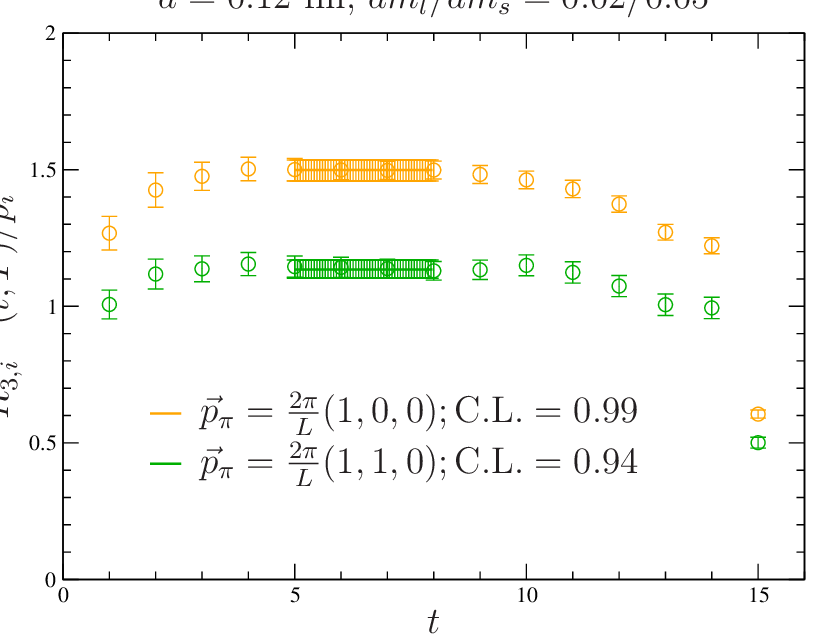}}
	\end{tabular}
	\caption{Determination of the form factors $f_\parallel$ (left plot) and $f_\perp$ (right plot) from plateau fits to the ratios defined in Eqs.~(\ref{eq:fpar_R}) and (\ref{eq:fperp_R}). The statistical errors on the data points are from a single-elimination jackknife.  The statistical errors in the plateau determination are from separate fits of 500 bootstrap ensembles.}
	\label{fig:R_plateaus}
\end{CFigure}

\subsection{Continuum and chiral extrapolation}
\label{sec:ChPT}

The quark masses in our numerical lattice simulations are heavier than
the physical up and down quark masses.  The effects of non-zero lattice
spacings in Asqtad simulations are also too large to be neglected.  In
order to account for these facts, we calculate the desired hadronic
matrix elements for multiple values of the light quark masses and lattice spacing, and then extrapolate to the physical quark masses and
continuum using functional forms from heavy-light meson staggered
chiral perturbation theory (HMS$\chi$PT)~\cite{Aubin:2007mc}.  The
HMS$\chi$PT expressions are derived using the symmetries of the
staggered lattice theory, and therefore contain the correct dependence
of the form factors on the quark mass and lattice spacing.  In the case
of the $B\to\pi\ell\nu$ form factors, the HMS$\chi$PT expressions are
also functions of the pion energy (recall that we work in the frame
where the $B$-meson is at rest).  

HMS$\chi$PT is a systematic expansion in inverse powers of the heavy
quark mass. In the chiral and soft pion limits ($m_l \to 0$ and $E_\pi
\to 0)$, the leading-order continuum HM$\chi$PT expressions for
$f_\parallel$ and $f_\perp$ take the following simple forms:
\begin{eqnarray}
     f_\parallel(E_\pi) & = & \frac{\phi_B}{f_\pi} \label{eq:fpar_LO} \\
     f_\perp (E_\pi) & = & \frac{\phi_{B^*}}{f_\pi} \frac{g_{B^* B \pi}}{E_\pi +
\Delta_B^*}, \label{eq:fperp_LO}
\end{eqnarray}
where $\phi_B \equiv f_B \sqrt{m_B}$, $f_B$ is the $B$-meson decay
constant, and $f_\pi$ is the pion decay constant.  The coefficient
$g_{B^* B \pi}$ parameterizes the size of the $B^*$-$B$-$\pi$ coupling.  In
the static heavy quark limit, heavy quark spin symmetry does not
distinguish between the pseudoscalar $B$-meson and the vector
$B^*$-meson, which implies that the decay constant $\phi_{B^*} = \phi_{B}$ and the mass difference $\Delta_B^* \equiv
m_{B^*} - m_B \to 0$.  Inclusion of the parameter $\Delta_B^*$,
however, ensures the proper location of the pole at $m_{B^*}^2$ in the
physical form factor $f_+(q^2)$.  At the next order in the heavy quark expansion,
$\CO(1/m_b)$ corrections split the degeneracy between the $B$- and
$B^*$-meson masses and decay constants.  Furthermore, in the chiral and soft pion limits, all $1/m_b$
corrections can be absorbed into the values of the parameters $\phi_B$,
$\phi_{B^*}$, $g_{B^* B \pi}$, and $\Delta_B^*$~\cite{Burdman:1993es};
thus $f_\parallel$ and $f_\perp$ retain the functional forms in Eqs.~(\ref{eq:fpar_LO}) and (\ref{eq:fperp_LO}) even at NLO in HM$\chi$PT.

At lowest-order in S$\chi$PT, discretization effects split the
degeneracies among the 16 tastes of pseudo-Goldstone mesons:
\begin{equation}
	m_{xy,\Xi}^2 = \mu \left( m_x + m_y \right) + a^2 \Delta_\Xi ,
\end{equation}
where $x$ and $y$ indicate the quark flavors, $\mu$ is a continuum
low-energy constant, and $\Delta_\Xi$ is the mass-splitting of a meson
with taste $\Xi$.  An exact $U(1)_A$ symmetry protects the taste
pseudoscalar meson from receiving a mass-shift to all orders in
S$\chi$PT, implying that $\Delta_P=0$.  In addition, at $\CO(a^2)$,  a
residual $SO(4)$ taste-symmetry preserves the degeneracies among mesons
that are in the same irreducible representation:  $P, V, A, T, I$~\cite{Lee:1999zx}.
Numerically, the size of the taste-splittings turn out to be comparable
to those of the pion masses for the $a=0.09$ fm and $a=0.12$ fm Asqtad
staggered lattices used in this work~\cite{Aubin:2004fs}.

We extrapolate our numerical form factor data using HMS$\chi$PT
expressions derived to zeroth order in $1/m_b$.
The fit functions therefore depend upon the three remaining expansion
parameters: $m_l$, $a$, and $E_\pi$.  The HMS$\chi$PT expressions for
the form factors to $\CO(m_l, a^2, E_\pi^2)$ are given explicitly in Eqs.(65)--(67) of Ref.~\cite{Aubin:2007mc}.  Schematically, they read
\begin{eqnarray}
	f_\parallel(m_l, E_\pi, a) & = &
\frac{c_\parallel^\textrm{(0)}}{f_\pi} \left[ 1 + \textrm{logs} +
c_\parallel^{(1)} m_l + c_\parallel^{(2)} (2m_l + m_s)  +
c_\parallel^{(3)} E_\pi + c_\parallel^{(4)} E_\pi^2 + c_\parallel^{(5)}
a^2 \right] \label{eq:fpar_ChPT}\\
	f_\perp(m_l, E_\pi, a) & = & \frac{c_\perp^\textrm{(0)}}{f_\pi} \left[
\frac{1}{E_\pi + \Delta_B^* + \textrm{logs}} + \frac{1}{E_\pi +
\Delta_B^*} \times \textrm{logs} \right] \nonumber\\
		& + &\frac{c_\perp^\textrm{(0)}/f_\pi}{E_\pi + \Delta_B^* } \left[
c_\perp^{(1)} m_l + c_\perp^{(2)} (2m_l + m_s) + c_\perp^{(3)} E_\pi +
c_\perp^{(4)} E_\pi^2 + c_\perp^{(5)} a^2 \right] ,
\label{eq:fperp_ChPT}
\end{eqnarray}
where ``logs" indicate non-analytic functions of the pseudo-Goldstone
meson masses, \emph{e.g.}, $m_\pi^2 \text{ln} (m_\pi^2 / \Lambda_\chi^2)$.  The
continuum low-energy constant $g_{B^* B \pi}$ enters these expressions in the
coefficients of the chiral logarithms, which are completely fixed at
this order.  We use the phenomenological value of $g_{B^* B \pi} = 0.51$~\cite{Arnesen:2005ez} for the central value and vary $g_{B^* B \pi}$ by a
reasonable amount (see Sec.~\ref{sec:gpi_err}) to estimate its
contribution to the systematic uncertainty.   Because the size of the
mass-splitting $\Delta^*_B$ is poorly determined from the lattice data
and is consistent with the physical value within statistical errors, we
fix $\Delta^*_B$ to the PDG value, 45.78 MeV~\cite{Amsler:2008zz}, in our
fits.  The chiral logarithms also depend upon six
extra constants that parameterize discretization effects due to the
light staggered quarks:  the four taste splittings $a^2 \Delta_V, a^2
\Delta_A, a^2 \Delta_T, a^2\Delta_I$ and the two flavor-neutral
``hairpin" coefficients $a^2 \delta'_V$ and $a^2
\delta'_A$~\cite{Aubin:2003mg}.  These parameters can be determined
separately from fits to light pseudoscalar meson masses and decay
constants;  we therefore hold them fixed to the values determined in Ref.~\cite{Bernard_PC} while performing the
continuum-chiral extrapolation.   The variation of these parameters
within their statistical errors results in a negligible change to the
extrapolated form factors.  The five terms analytic in $m_l$, $a^2$,
and $E_\pi$ absorb the dependence upon the scale in the chiral
logarithms, $\Lambda_\chi$, such that the form factor is
scale-independent.   We leave the tree-level coefficients
$c^{(0)}_{\parallel,\perp}$ and the NLO analytic term coefficients
$c^{(1)}_{\parallel,\perp}$--$c^{(5)}_{\parallel,\perp}$ as free
parameters to be determined via the fit to the lattice form
factor data.  In practice, we omit the analytic term proportional to
$(2m_l + m_s)$ from our fits because the strange sea quark mass is tuned to approximately the same value on each of our ensembles and we have simulated only full QCD points.  This
term is therefore largely indistinguishable from the analytic term
proportional to $m_l$.  We have checked that omission of the sea
quark mass analytic term has a negligible impact on the form
factors in the chiral and continuum limits.

\bigskip

In both earlier unquenched analyses of the $B\to\pi\ell\nu$
semileptonic form factor~\cite{Okamoto:2004xg,Dalgic:2006dt}, the chiral extrapolation is
performed as a two-step procedure:  first interpolate the lattice data
to fiducial values of $E_\pi$ and then extrapolate the results to the
physical quark masses and continuum independently at each value of
$E_\pi$.  The function used for the interpolation (which is different in the two analyses) introduces a systematic uncertainty that is difficult to estimate.  In both cases, 
the chiral-continuum extrapolation makes use of the correct functional
forms derived in HMS$\chi$PT, but, by extrapolating the results for each value of
$E_\pi$ separately, the constraint that
the low-energy constants of the chiral effective Lagrangian are
independent of the pion energy is lost.
This omission of valuable information about the form factor shape introduces a further error that is unnecessary.  The new analysis presented here instead employs a simultaneous
fit using HMS$\chi$PT to our entire data set (all values of $m_l$, $a$, and $E_\pi$) to extrapolate to physical quark masses and the continuum and
interpolate in the pion energy~\cite{Aoki:2001rd}.  This improved method eliminates the systematic
uncertainties introduced in the two-step interpolate-then-extrapolate procedure, and exploits the available information in an optimal way.

We perform our combined chiral and continuum extrapolation using the
method of constrained curve fitting~\cite{Lepage:2001ym}.  Although we
know that lattice data generated with sufficiently small quark masses
and fine lattice spacings, and, in the case of the $B\to\pi\ell\nu$
form factor, sufficiently low pion energies, must be described by
lattice $\chi$PT, we do not know precisely the range of validity of the
effective theory.  Furthermore, the order in $\chi$PT to which we must
work and the allowed parameter values depend upon both the quantity of
interest and the size of the statistical errors.  We therefore need a fitting procedure that both incorporates our general theoretical understanding of the suitable chiral effective
theory and accounts for our limited
knowledge of the values of the low-energy constants and sizes of the
higher-order terms.  Constrained curve fitting provides just such a
method.  

Next-to-leading order $\chi$PT breaks down for pion energies around and above the kaon mass.
 Less than half of our numerical form factor data, however,
is below this cutoff.  Therefore, although we do not
expect NLO HMS$\chi$PT to describe our data through momentum $p =
2\pi(1,1,0)/L$, we cannot remove those points without
losing the majority of our data.  Nor can we abandon the NLO
HMS$\chi$PT expressions for $f_\parallel$ and $f_\perp$,
Eqs.~(\ref{eq:fpar_ChPT}) and~(\ref{eq:fperp_ChPT}), which are the only
effective field theory guides that we have for extrapolating the numerical lattice form factor
data to the continuum and physical quark masses.  We therefore perform the continuum-chiral
extrapolation using the full NLO HMS$\chi$PT expressions for
$f_\parallel$ and $f_\perp$, including the 1-loop chiral logarithms,
\emph{plus} additional NNLO analytic terms to allow a good fit to the
data through $p = 2\pi(1,1,0)/L$.  The NNLO terms smoothly interpolate
between the region in which $\chi$PT is valid and the region in which
the pion energies are too large and the higher-order chiral logarithms
in $E_\pi$ can be approximated as polynomials.

We express the analytic terms in the formulae for $f_\parallel$ and
$f_\perp$, Eqs.~(\ref{eq:fpar_ChPT}) and~(\ref{eq:fperp_ChPT}), as
products of dimensionless expansion parameters:
\begin{eqnarray}
	\chi_{m_l} & = & \frac{2 \mu m_l}{8 \pi^2 f_\pi^2} \sim 0.05 \text{--} 0.19 \\
	\chi_{a^2} & = & \frac{a^2 \bar\Delta}{8 \pi^2 f_\pi^2} \sim 0.03 \text{--} 0.09 \\
	\chi_{E_\pi} & = & \frac{\sqrt{2} E_\pi}{4\pi f_\pi} \sim 0.22  \text{--} 0.78 ,
\end{eqnarray}
where $\bar{\Delta}$ is the average staggered taste-splitting and we show the range of values for each of these parameters
corresponding to our numerical lattice data.  (Note that we omit the
$\vec{p}=2\pi(1,1,1)/L$ data points from our chiral fits because these would
lead to $\chi_{E_\pi} \gtapprox 1$.)  Because each of the above
expressions is normalized by the chiral scale, $\Lambda_\chi \approx 4
\pi f_\pi$, the undetermined coefficients
$c^{(1)}_{\parallel,\perp}$--$c^{(5)}_{\parallel,\perp}$ should be of
$\CO(1)$ in these units.  We therefore constrain the values of the
low-energy constants
$c^{(0)}_{\parallel,\perp}$--$c^{(5)}_{\parallel,\perp}$ in our fits
with Gaussian priors of width 2 centered about 0.

The statistical errors in the numerical lattice data come from the
3-point fits described in the previous subsection.  In order to account
for the correlations among the various pion energies on the same sea
quark ensemble in the chiral-continuum extrapolation, we preserve the
bootstrap distributions.  We perform a separate correlated fit to each
of the 500 bootstrap ensembles in which we remake the full bootstrap
covariance matrix for each fit.  We average the $68\%$ upper and lower
bounds on the form factor distributions to determine the statistical and systematic errors in $f_\parallel$ and $f_\perp$ that are plotted in Fig.~\ref{fig:ChPT_extrap} and presented in Table~\ref{tab:fP_err} below.

Because we do not know \emph{a priori} how many terms are necessary to
describe the available lattice data, we begin with strictly NLO fits
using the formulae for $f_\parallel$ and $f_\perp$ in
Eqs.~(\ref{eq:fpar_ChPT}) and~(\ref{eq:fperp_ChPT}).  We fit the
lattice data for $f_\parallel$ and $f_\perp$ separately even though the
ratio of leading-order coefficients, $c^{(0)}_{\perp} /
c^{(0)}_{\parallel}$, is predicted to equal $g_{B^* B \pi}$ to NLO in $\chi$PT;
this is because the value of $g_{B^* B \pi}$ is known to only $\sim 50\%$ from
phenomenology.  We obtain a good fit of the $f_\perp$ lattice data to
the NLO expression without the inclusion of higher-order NNLO terms.
This is probably because the shape of $f_\perp$ is dominated by the
$1/(E_\pi + \Delta^*)$ behavior and therefore largely insensitive to
the other terms.  We cannot, however, obtain a good fit of
$f_\parallel$ to the strictly NLO expression, and must add higher-order
terms in order to obtain a successful fit.  Specifically, NNLO analytic terms proportional to $m_l E_\pi$
and $E_\pi^3$ are both necessary to achieve a confidence level better
than $10\%$.  

Although we could, at this point, choose to truncate the HMS$\chi$PT
extrapolation formulae to include only those terms necessary for a good
confidence level, we instead include ``extra" NNLO analytic terms to
both the $f_\parallel$ and $f_\perp$ fits, constraining the values of
their coefficients with Gaussian priors of $0 \pm 2$.  The introduction
of more free parameters increases the statistical errors in the
extrapolated values of the form factors; these larger errors reflect
the uncertainty in the size of the newly-included higher-order
contributions. We continue to add higher-order analytic terms until the
central values of the extrapolated form factors stabilize and the
statistical errors in the form factors reach a maximum.  This indicates
that any further terms are of sufficiently high order that they do not
affect the fit and can safely be neglected.  We find that this occurs
once the extrapolation formulae for $f_\parallel$ and $f_\perp$ contain
all eight sea-quark mass-independent NNLO analytic terms.  The
inclusion of NNNLO analytic terms does not further increase the size of
the error bars.

Figure~\ref{fig:ChPT_extrap} shows the preferred constrained fits of
$f_\parallel$ (upper plot) and $f_\perp$ (lower plot) versus $E_\pi^2$,
where both the $x$- and $y$-axes are in $r_1$ units.\footnote{As a cross-check of the constrained fits, we also perform unconstrained fits of $f_\parallel$ and $f_\perp$ with only the minimal
number of analytic terms needed for a good fit.  The results are consistent, but the unconstrained fit results have smaller statistical errors because they include 6--8 fewer fit
parameters.}  Each fit is to the NLO HMS$\chi$PT expression,
Eqs.~(\ref{eq:fpar_ChPT}) and~(\ref{eq:fperp_ChPT}), plus all sea-quark
mass-independent NNLO analytic terms.  The square symbols indicate fine
lattice data, while the circles denote coarse data.  The six colored
curves show the fit result projected onto the masses and lattice
spacings of the six sea quark ensembles; the red line should go through
the red circles, and so forth.  The thick black curve shows the form
factor in the continuum at physical quark masses with symmetrized bootstrap
statistical errors.

\begin{CFigure}
\rotatebox{0}{\includegraphics[width=0.7\textwidth]{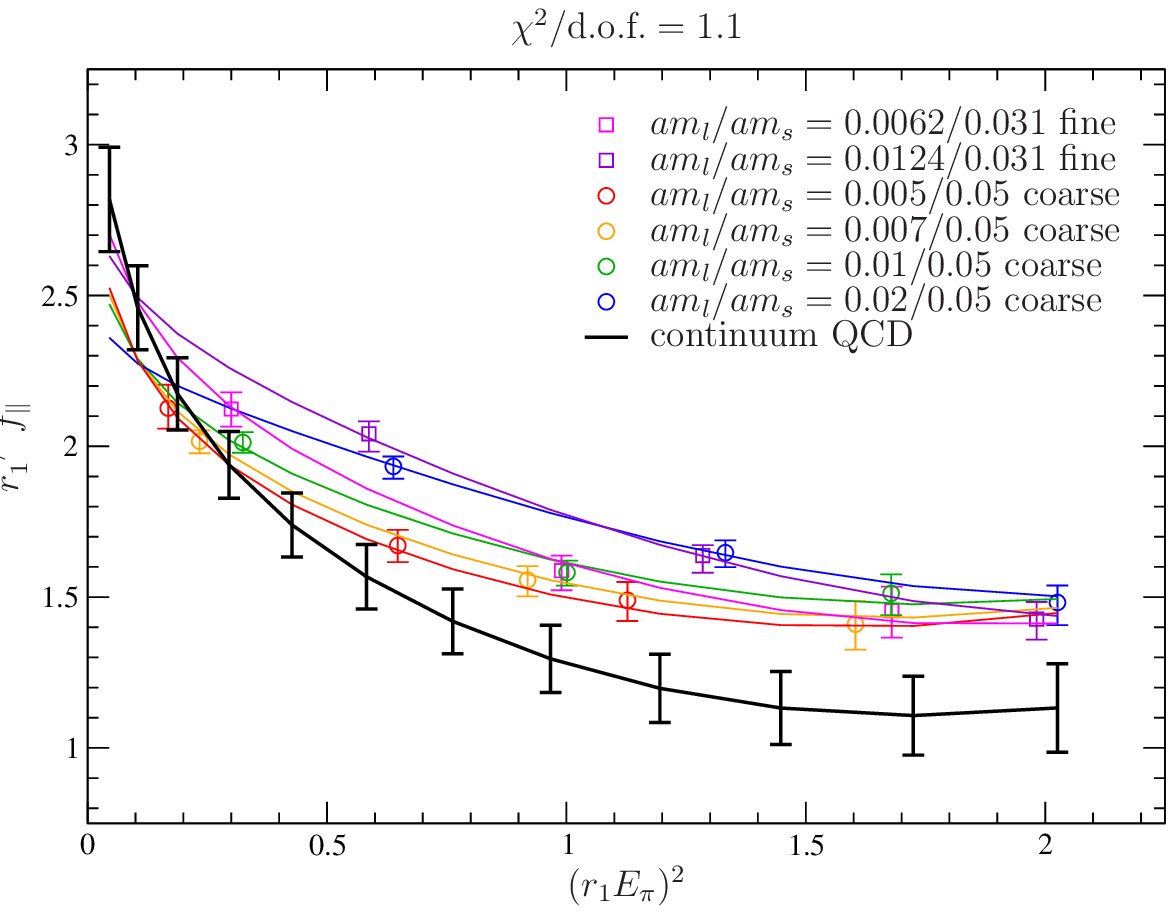}}
\rotatebox{0}{\includegraphics[width=0.7\textwidth]{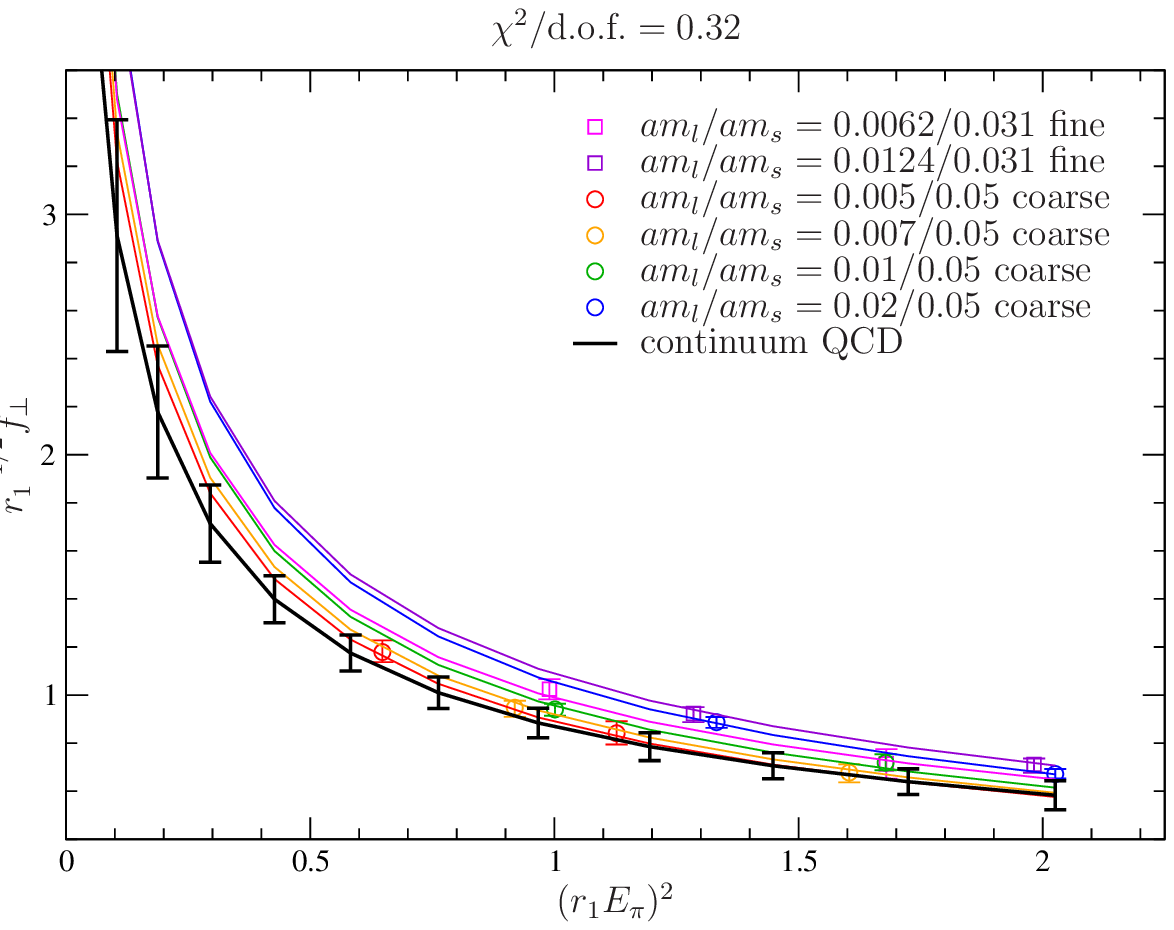}}
	\caption{Chiral-continuum extrapolation of $f_\parallel$ (upper) and
$f_\perp$ (lower) using constrained NLO HMS$\chi$PT plus all NNLO
analytic terms with $g_{B^* B \pi}=0.51$ and $r_1=0.311$ fm.  The square
symbols indicate $a \approx 0.09$ fm lattice data points while the circular
symbols indicate $a \approx 0.12$ fm coarse data points.  The black curve is the
chiral-continuum extrapolated form factor symmetrized bootstrap statistical errors only.}\label{fig:ChPT_extrap}
\end{CFigure}  

We use functions and constraints based on HMS$\chi$PT to perform the
chiral-continuum extrapolation because we know that HMS$\chi$PT is the
correct low-energy effective description of the lattice theory.
Nevertheless, we must compare various properties with theoretical
expectations in order to check for overall consistency.  An essential
first test is that we can successfully fit the data with good
confidence levels and obtain low-energy coefficients that are of the
predicted size.  We can also verify the convergence of the series
expansion by calculating the ratios of the higher-order contributions
to the leading-order form factor contributions:
\begin{eqnarray} 
	\frac{f_\parallel^\text{NLO}}{f_\parallel^\text{LO}}\bigg|_{E_\pi =
500 \text{ MeV}}& \approx &47\% ,\qquad
	\frac{f_\perp^\text{NLO}}{f_\perp^\text{LO}}\bigg|_{E_\pi = 500 \text{
MeV}} \approx 48\% ,\\
	\frac{f_\parallel^\text{NNLO}}{f_\parallel^\text{LO}}\bigg|_{E_\pi =
500 \text{ MeV}} &\approx& 3\% ,\qquad
	\frac{f_\perp^\text{NNLO}}{f_\perp^\text{LO}}\bigg|_{E_\pi = 500
\text{ MeV}} \approx 4\% ,
\end{eqnarray}
where we choose a nominal value of $E_\pi = 500 \text{ MeV}$ for
illustration because it is on the high end of the expected range of
validity of $\chi$PT.  Finally, because the leading-order coefficient,
$c^{(0)}_{\parallel}$, is expected to be equal to $\phi_B \equiv f_B
\sqrt{m_B}$ in HM$\chi$PT, we can compare its value with that of $\phi_B$
determined from our preliminary decay
constant analysis.  Although the $B$-meson decay constant calculation
uses the same staggered gauge configurations, it employs different
heavy-light meson 2-point correlation functions with the axial current,
a different HMS$\chi$PT fit function, and different perturbative
renormalization factors, and is therefore largely independent of the
$B\to\pi\ell\nu$ semileptonic form factor calculation.  For the
preferred $f_\parallel$ fit shown in Fig.~\ref{fig:ChPT_extrap}, we
find $c^{(0)}_{\parallel} = 0.81\pm 0.07$, where the errors are
statistical only.  This is quite close to our current preliminary
result, $r_1^{3/2} \phi_B = 0.92 \pm  0.03$
(statistical error only)~\cite{Bernard:2007zz,Bernard_fB}, especially
considering that the HMS$\chi$PT extrapolation formula for $f_\parallel$
neglects some of the $\CO(1/m_b)$ contributions.

An interesting use of our numerical $B\to\pi\ell\nu$ form factor data
is to determine the approximate value of the $B^*$-$B$-$\pi$ coupling,
$g_{B^* B \pi}$, from lattice QCD.  For the preferred fits shown in
Fig.~\ref{fig:ChPT_extrap}, we find that the ratio of leading-order
coefficients is
\begin{equation}
 	g_{B^* B \pi} \approx \frac{c^{(0)}_{\perp}}{ c^{(0)}_{\parallel}} =
0.22 \pm 0.07 ,
 \end{equation} 
and is independent of the choice for $g_{B^* B \pi}$ in the chiral logarithms
within statistical errors.  This determination omits the $\CO(1/m_b)$ corrections to the chiral logarithms
in the HMS$\chi$PT extrapolation formulae for $f_\parallel$ and
$f_\perp$, Eqs.~(\ref{eq:fpar_ChPT}) and~(\ref{eq:fperp_ChPT}), and
neglects the difference between $\phi_B$ and $\phi_{B^*}$; we do not
attempt to estimate the systematic uncertainty introduced by these or
other effects.  The value is lower than the determination of Stewart,
$g_{B^* B \pi} = 0.51$, which comes from a combined analysis of several experimental
quantities, including the $D^*$-meson decay width, through $\CO(1/m_c)$ in
HM$\chi$PT~\cite{Arnesen:2005ez}.   It is consistent, however, with the
range of values determined by the HPQCD Collaboration, who allowed
$g_{B^* B \pi}$ to be a free parameter in their chiral-continuum extrapolation
and found $0 < g_{B^* B \pi} \ltapprox 0.45$~\cite{Dalgic:2006dt}.  

\section{Estimation of systematic errors}
\label{sec:Errors}

In this section, we discuss all of the sources of systematic uncertainty
in our calculation of the $B\to\pi\ell\nu$ form factor $f_+(q^2)$.  We
present each error in a separate subsection for clarity.  The value of the form factor $f_+(q^2)$, along with the total error budget, is given in Table~\ref{tab:fP_err}.

\subsection{Chiral-continuum extrapolation fit ansatz}

We use the method of constrained curve fitting to estimate the effect
of neglected higher-order terms in the HMS$\chi$PT chiral-continuum
extrapolation formulae.  Our fit procedure is described in detail in
Sec.~\ref{sec:ChPT}.  Therefore, the errors in $f_\parallel$ and
$f_\perp$ extrapolated to physical quark masses and zero lattice
spacing shown in Fig.~\ref{fig:ChPT_extrap} reflect \emph{both} the
statistical errors in the Monte Carlo data and the systematic errors
due to our limited knowledge of the higher-order terms, which we
specified with priors.  We do not need to include a separate systematic
uncertainty due to the choice of fit function, as would be the case had
we used an unconstrained fit with fewer terms.

\subsection{\boldmath $g_{B^* B \pi}$ uncertainty}
\label{sec:gpi_err}

We fix the size of the $B^*$-$B$-$\pi$ coupling to $g_{B^* B \pi}=0.51$ in the
coefficients of the chiral logarithms while extrapolating to the
physical light quark masses and continuum.  Our choice is based upon the following considerations.  Because the coupling $g_{B^*B\pi}$ is expected to be approximately equal to $ g_{D^*D\pi}$ due to heavy-quark symmetry, we use the phenomenological value of the $D^*$-$D$-$\pi$ coupling determined by Stewart in Ref.~\cite{Arnesen:2005ez}, which comes from a combined analysis of several experimental quantities that includes the $D^*$-meson decay width~\cite{Anastassov:2001cw}.  This value is presented without errors, and is an update of
Stewart's earlier analysis in Ref.~\cite{Stewart:1998ke} which incorporates additional experimental results.  His earlier calculation finds a significantly lower value of $g_{D^*D\pi}=0.27^{+0.04\, + 0.05}_{-0.02\, -0.02}$, where the first errors are experimental and the second errors come from an estimate of the sizes of the 1-loop counterterms~\cite{Stewart:1998ke}.  A more recent phenomenological determination of the $D^*$-$D$-$\pi$ coupling by Kamenik and Fajfer, which also includes up-to-date experimental data, improves upon the analysis method of Stewart by including contributions from both positive and negative parity heavy mesons in the loops~\cite{Fajfer:2006hi}.  They find an even higher value of $g_{D^*D\pi}=0.66^{+0.08}_{-0.06}$, where the uncertainty only reflects the error due to counterterms.  We therefore conclude that, although recent experimental measurements of the $D^*$ width may constrain the coupling~\cite{Anastassov:2001cw} at tree-level, the size of $g_{B^* B \pi}$ is not well-determined in the literature.  

In order to determine the error in the form factor from the uncertainty
$g_{B^* B \pi}$ we vary the parameter over a generous range.  The smallest
value of $g_{B^* B \pi}$ that we have seen in the literature is $g_{D^*D\pi} =
0.27$~\cite{Stewart:1998ke}.  The largest is $g_{D^*D\pi}=0.67$, which
comes from a quenched lattice calculation~\cite{Abada:2002vj}.  (There
has not yet been an unquenched ``2+1" flavor determination of $g_{B^* B \pi}$.)
We therefore vary $g_{B^* B \pi}$ over the entire range from 0.27--0.67 and
take the largest difference from the preferred determination of
$f_+(q^2)$ using $g_{B^* B \pi} = 0.51$ as the systematic error due to the
uncertainty in the $B^*$-$B$-$\pi$ coupling.   The lattice
data is largely insensitive to the value of $g_{B^* B \pi}$ in the coefficient
of the chiral logarithms; all values of the parameter yield similar
fit confidence levels. The resulting systematic uncertainty in
$f_+(q^2)$ is less than 3\% for all $q^2$ bins despite varying $g_{B^* B \pi}$
by almost 50\%.

\subsection{\boldmath Scale ($r_1$) uncertainty}

We use the MILC Collaboration's determination of the scale from their
calculation of $f_\pi$, $r_1=0.311$ fm, to convert between lattice and
physical units~\cite{Bernard:2007ps}.  The parameter $r_1$ enters the
form factor calculation in a number of places:  we use the PDG values
of $f_\pi$ and $\Delta^*_B$ in the chiral-continuum
extrapolation formulae~\cite{Amsler:2008zz}, we set $m_\pi$ to the PDG value in the
resulting fit functions to determine the form factors at the physical
point, and we convert $f_\parallel$ and $f_\perp$ to physical units via
$r_1$ before combining them to extract $f_+(q^2)$.  An alternative
determination using the HPQCD Collaboration's lattice data for the $\Upsilon$ $2S$-$1S$~\cite{Gray:2005ur} splitting yields a result  that is $\sim$ 2\%
larger, $r_1=0.317$ fm.  We therefore repeat the chiral-continuum
extrapolation of $f_\parallel$ and $f_\perp$ using this higher value of
$r_1$, combine them into the dimensionless form factor $f_+(q^2)$ using
this higher value of $r_1$, and take the difference from the preferred
form factor result as the systematic error due to uncertainty in the
overall lattice scale.  The difference ranges from 1--1.5\% for most
$q^2$ values.  This is consistent with our naive expectation that a
$\sim 2\%$ difference in $r_1$ will result in a $\sim 1\%$ difference
in $f_+(q^2)$ because $f_\parallel$ has dimensions of
${\text{GeV}}^{1/2}$ and $f_\perp$ has dimensions of
${\text{GeV}}^{-1/2}$.

\subsection{\boldmath Light quark mass ($\hat{m}$, $m_s$) determinations}

We obtain the form factors $f_\parallel$ and $f_\perp$ in continuum QCD
by setting the lattice spacing to zero and the light quark masses to
their physical values in the HMS$\chi$PT expressions, once the
coefficients have been determined from fits to the numerical lattice data.  We use
the most recent calculations of the bare quark masses by the MILC
Collaboration from fits to light pseudoscalar meson masses:
\begin{eqnarray}
	r_1 \hat{m} \times 10^3 & = & 3.78 (16) \\
	r_1 m_s \times 10^3 & = & 102 (4),
\end{eqnarray}
where $\hat{m}$ is the average of the up and down quark masses and the quoted errors
include both statistics and systematics~\cite{Bernard:2007ps}.
We vary the bare
light quark mass, $r_1 \hat{m}$, within its stated uncertainty and take
the maximal difference from the preferred form factor result to be the
systematic error; we find that the error is $0.3\%$ or less for all
values of $q^2$.  We perform the same procedure for the bare strange
quark mass, and find that the resulting error ranges from $\sim$
0.5--1.5\% over the various $q^2$ bins.

\subsection{\boldmath Bottom quark mass ($m_b$) determination}
\label{sec:kappab}

The value of the form factor $f_+(q^2)$ depends upon the $b$-quark
mass, which we fix to its physical value throughout the calculation.
Specifically, we first determine the value of the hopping parameter,
$\kappa$, in the SW action for which the lattice kinetic mass
agrees with the experimentally-measured $B_s$-meson mass.  We then use
this tuned $\kappa_b$ when calculating all of the 2- and 3-point
heavy-light correlators needed for the $B\to\pi\ell\nu$ form factor.
With our current tuning procedure we are able to determine $\kappa_b$
to $\sim 6\%$ accuracy.  This uncertainty in $\kappa_b$ is
conservative;  it is primarily due to poor statistics, and will
decrease considerably after the analysis of the larger data set that is
currently being generated.

The uncertainty in $\kappa_b$ produces an uncertainty in the form
factor.  We estimate this by calculating the form factor $f_+$ at two additional values of $\kappa$ (one above and one below the tuned value) on the $a m_l / a m_s = 0.02/0.05$ coarse ensemble.  This is sufficient because the
heavy-quark mass-dependence of the form factor is largely independent
of the sea quark masses and lattice spacing.   We find the largest
dependence upon $\kappa_b$ at momentum
$\vec{p}=2 \pi (1,1,0) / L$, shown in Fig.~\ref{fig:dff_dk}, for which a 6\%
uncertainty in $\kappa_b$ produces a 1.2\% uncertainty in the form
factor.  We therefore take 1.2\% to be the systematic error in
$f_+(q^2)$ due to uncertainty in the determination of the $b$-quark
mass.

\begin{CFigure}
\rotatebox{0}{\includegraphics[width=0.55\textwidth]{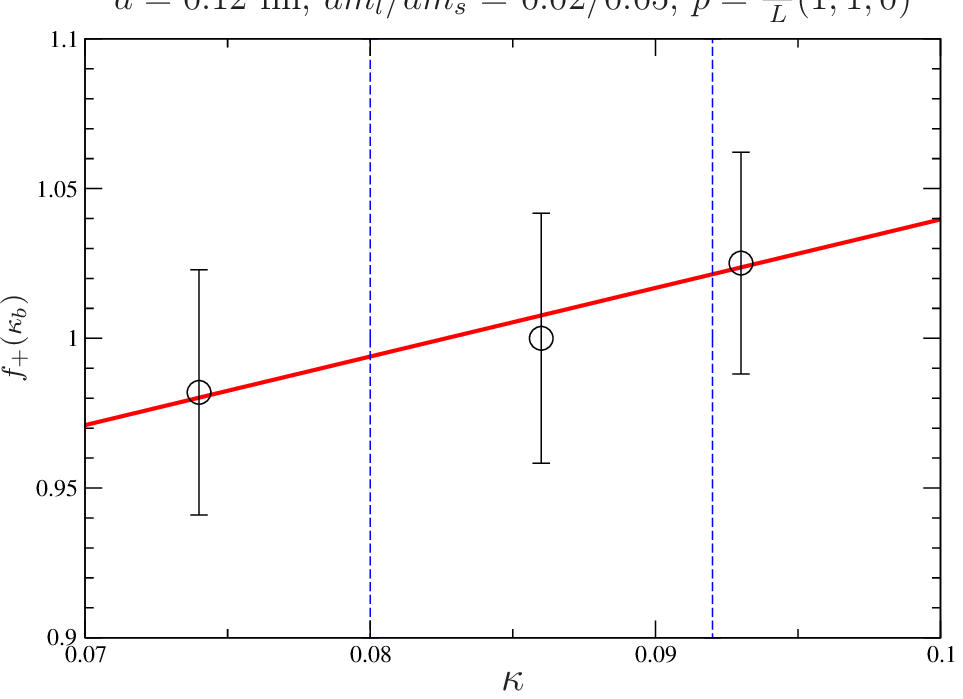}}
	\caption{Normalized form factor $f_+$ at momentum
$\vec{p}=2 \pi (1,1,0) / L$ as a function of $\kappa$ on the
$am_l/am_s=0.02/0.05$ coarse ensemble.  The central data point
corresponds to the tuned $\kappa_b$, and the thick red line shows a
linear fit to the three data points.  The two dashed vertical lines
indicate the upper and lower bounds on $\kappa_b$.}\label{fig:dff_dk}
\end{CFigure}  

\subsection{Gluon and light-quark discretization errors}

We estimate the size of discretization errors in the form factor
$f_+(q^2)$ with power-counting.  We choose conservative values for the
parameters that enter the estimates: $\Lambda = 700$ MeV and $\alpha_V(q^*) =
1/3$, which is a typical value on the fine lattice spacing~\cite{ElKhadra:2007qe,ElKhadra_PT}.

We calculate the $B\to\pi\ell\nu$ semileptonic form factor using a
one-loop Symanzik-improved gauge action for the gluons
\cite{Alford:1995hw,Bernard:1997mz,Luscher:1984xn,Luscher:1985zq} and
the Asqtad-improved staggered action for the light up, down, and
strange quarks \cite{Lepage:1998vj,Orginos:1999cr}.   Because both the gluon and
light quark actions are $\CO(a^2)$-improved, the leading discretization
effects are  of $\CO(\alpha_s (a \Lambda)^2)$.  We parameterize these
errors in the fit to numerical lattice form factor data by including
analytic terms proportional to $a^2$ in the HMS$\chi$PT extrapolation
formulae for $f_\parallel$ and $f_\perp$.  Because we only have data at two lattice spacings, however, we do not include a separate term proportional to $\alpha_s (a \Lambda)^2$ to account for the fact that $\alpha_s$ differs by a few percent between the lattice spacings.   We then remove the majority of light quark and gluon discretization effects from the final result by taking $a\to 0$.  Similarly, we identify and remove higher-order discretization effects in the chiral-continuum
extrapolation through the NNLO analytic terms in the fit functions.
The remaining gluon and light quark discretization errors are
negligible. 

The calculation of the $B\to\pi\ell\nu$ form factor requires 2-point
and 3-point functions with nonzero momenta, which introduces
momentum-dependent discretization errors.  The leading $p$-dependent
discretization error is of $\CO(\alpha_s (ap)^2)$.  We parameterize
these errors, up to variations in $\alpha_s$ at the two lattice spacings, with the two NNLO analytic terms proportional to $a^2 E_\pi^2$ and $a^2 m_l$ in the extrapolation formulae for $f_\parallel$
and $f_\perp$ and remove them by taking the continuum limit of the
resulting fit functions.  This also largely removes errors of $\CO(\alpha_s^2 (ap)^2)$.  The remaining momentum-dependent discretization effects are of $\CO(\alpha_s
(ap)^4)$.  On the $28^3$ fine lattices, 
$\alpha_s (ap)^4 = 0.003$ for our highest-momentum data points with
$a\vec{p}=2\pi(1,1,0)/28$. Therefore the uncertainty in $f_+(q^2)$ due to momentum-dependent discretization
effects is negligible compared with our other systematic errors.

\subsection{Heavy-quark discretization errors}

We use HQET as a theory of cutoff effects to estimate the size of discretization errors due to use of the Fermilab action for the heavy  bottom quark.  Because both the lattice and continuum theories can be described by HQET,  heavy-quark discretization effects can be classified as a short-distance mismatch 
of higher-dimension operators~\cite{Kronfeld:2000ck,Harada:2001fi,Harada:2001fj}.
Each contribution to the error is given by~\cite{Kronfeld:2003sd}
\begin{equation}
	{\tt error}_i = \left|\left[
		{\cal C}_i^{\rm lat}(m_Q, m_0a) - {\cal C}_i^{\rm cont}(m_Q)
		\right] \langle{\cal O}_i\rangle \right|,
\end{equation}
where ${\cal O}_i$ is an effective operator, and
${\cal C}_i^{\rm lat}(m_Q, m_0a)$ and ${\cal C}_i^{\rm cont}(m_Q)$ are 
the corresponding short-distance coefficients when HQET is used to 
describe lattice gauge theory or continuum QCD, respectively.
The coefficient mismatch can be written as
\begin{equation}
	{\cal C}_i^{\rm lat}(m_Q, m_0a) - {\cal C}_i^{\rm cont}(m_Q) =
		a^{\dim{\cal O}_i-4} f_i(m_0a),
\end{equation}
and the relative error in our matrix elements can be estimated by setting
$\langle{\cal O}_i\rangle\sim\Lambda_{\text{QCD}}^{\dim{\cal O}_i-4}$.
Then each contribution to the error is
\begin{equation}
	{\tt error}_i = f_i(m_0a) (a\Lambda_{\text{QCD}})^{\dim{\cal O}_i-4},
	\label{eq:error}
\end{equation}
recovering the counting in powers of $a$ familiar from Symanzik, while
maintaining the full $m_0a$ dependence.
The functions~$f_i$ can be deduced from
Refs.~\cite{ElKhadra:1996mp,Oktay:2008ex} and are compiled in 
Appendix~\ref{sec:App}.
Adding all contributions of $\CO(\alpha_sa)$ and $\CO(a^2)$ from the action 
and the current, we obtain a relative error of
2.84\% (4.16\%) for $f_\parallel$ and
3.40\% (4.98\%) for $f_\perp$ on the fine (coarse) lattices.
We therefore take 3.4\% to be the error in $f_+(q^2)$ due to heavy-quark discretization effects.

\subsection{Heavy-light current renormalization}

We determine the majority of the heavy-light current renormalization
nonperturbatively.  The dependence of $Z_V^{bb}$ on the sea quark
masses and on the mass of the light spectator quark in the 3-point
correlator are both negligible; the statistical error in $Z_V^{bb}$ is
$\sim 1\%$.  The dependence of $Z_V^{ll}$ on the sea quark masses is
also negligible, and the statistical error in $Z_V^{ll}$ is $\sim 1\%$.
We therefore include $\sqrt{(1\%^2 + 1\%^2}) \approx 1.4\%$ as the
systematic uncertainty in the form factor $f_+(q^2)$ due to the
uncertainty in the nonperturbative renormalization factors $Z_V^{bb}$
and  $Z_V^{ll}$ for all values of $q^2$.

We determine the remaining renormalization of the heavy-light current
using lattice perturbation theory.    The 1-loop correction to
$f_\perp$ is $\sim 3\%$ on the fine ensembles and $\sim 4\%$ on the
coarse ensembles.  Because we calculate $\rho^{hl}_{V_\mu}$ to
$\CO(\alpha_s)$, the leading corrections are of $\CO(\alpha_s^2)$.  We
might therefore expect the 2-loop corrections to $\rho^{hl}_{V_\mu}$ to be a factor
of $\alpha_s$ smaller, or $\sim 1\%$.  In order to be conservative, however, we take the entire size of the 1-loop correction on the fine lattices, or 3\%, as the systematic uncertainty in $f_+(q^2)$ due
to higher-order perturbative contributions for all $q^2$ bins.

\subsection{\boldmath Tadpole parameter ($u_0$) tuning}


In order to improve the convergence of lattice perturbation theory, we use tadpole-improved actions for the gluons, light quarks, and heavy quarks~\cite{Lepage:1992xa}.  We take $u_0$ from the average plaquette for the gluon and sea quark action~\cite{Bernard:2001av}.  On the fine lattice, we make the same choice for the valence quarks.  For historical reasons, however, we use $u_0$ determined from the average link in Landau gauge for the valence quarks on the coarse ensembles.  The difference between $u_0$ from the two methods is 3--4\% on the coarse ensembles.  We must, therefore,
estimate the error in the form factor due to this poor choice of tuning.  

The tadpole-improvement factor enters the calculation of $f_+(q^2)$ in several ways.  The factor of $u_0$ that enters the normalization of the heavy Wilson and
light staggered quark fields cancels exactly between the
$\langle \pi | V^{\mu} | B \rangle$ lattice matrix element and the
nonperturbative renormalization factor $\sqrt{Z_V^{bb} Z_V^{ll}}$.  The most significant effect of the mixed $u_0$ values is in the chiral-continuum extrapolation of $f_\parallel$ and
$f_\perp$.  The different choices for valence and sea quark actions imply that the coarse lattice data is partially quenched.  We study this effect by performing the chiral extrapolation in two ways: one assuming that both valence and sea quarks have the mass of the sea quark and the other assuming that both have the mass of the valence quark.  This leads to a 3\% error in the highest $q^2$ bin, and a $\sim 1 - 1.5\%$ error in the bins that affect the determination of~$|V_{ub}|$.  Most of the other effects of
changing $u_0$ in the lattice action and current can be absorbed into our estimate of the uncertainty from higher-order perturbative corrections to $\rho_V^{hl}$, to discretization errors, and to the normalization of the Naik term.  All but the last are already budgeted in Table~\ref{tab:fP_err}.  The Naik term in the Asqtad action ensures that the leading
discretization errors in the pion dispersion relation are $\CO(\alpha_s a^2
p^2)$.  We therefore estimate the error in $f_+(q^2)$ due to different
Naik terms in the valence and sea sectors to be equal to the largest
value of $\alpha_s a^2 p^2$ on the coarse lattice times the ratio of the Landau
link over plaquette $u_0$ cubed, or $\sim 0.2\%$. 

We add the flat error from the Naik term to the bin-by-bin error due to the light quark mass used in the chiral extrapolation in quadrature to obtain the total uncertainty.  Although our estimate is of necessity rather rough, we find that the errors due to $u_0$ tuning are much smaller than the dominant errors in $f_+(q^2)$.  Our error estimate is therefore adequate for the determinations of the $B\to\pi\ell\nu$ form factor and $|V_{ub}|$ presented in this work.

\subsection{Finite volume effects}

We estimate the uncertainty in the form factor $f_+(q^2)$ due to
finite-volume effects using 1-loop finite volume HMS$\chi$PT.  
The finite volume corrections to the HMS$\chi$PT expressions for $f_\parallel$ and $f_\perp$
are given in Ref.~\cite{Aubin:2007mc} in terms of integrals calculated in Ref.~\cite{Arndt:2004bg}.
It is therefore straightforward to find the
relevant corrections for our simulation parameters.  We find that the 1-loop finite volume corrections are well below a percent for all of our lattice data points.  Because finite
volume errors increase as the light quark mass decreases, they are
largest on the $am_l/am_s = 0.007/0.05$ coarse ensemble.  The biggest
correction is to $f_\perp$ at $\vec{p}=2 \pi (1,1,0) / L$, and is $0.5\%$.  We
therefore take this to be the uncertainty in $f_+(q^2)$ due to finite
volume errors for all $q^2$ bins.

\begin{sidewaystable}
\caption{Statistical and systematic error contributions to the
$B\to\pi\ell\nu$ form factor.  Each source of uncertainty is
discussed in Sec.~\ref{sec:Errors}.  For each of the 12 $q^2$ bins, the
error is shown as a percentage of the total form factor, $f_+(q^2)$, which is given in the second row from the top.  Because the bootstrap errors in the form factor are asymmetric, the errors shown are
the average of the upper and lower bootstrap errors.  In order to facilitate the use of our result, we also present the normalized statistical and systematic bootstrap correlation matrices in Table~\ref{tab:fP_cor} and the total bootstrap covariance matrix in Table~\ref{tab:fP_tot_cor}.
\medskip}
\label{tab:fP_err}
\begin{tabular}{lrrrrrrrrrrrr}
\hline\hline
$q^2$ (GeV$^2$)  & 26.5  & 25.7  & 25.0  & 24.3  & 23.5  & 22.8  & 22.1  & 21.3  & 20.6  & 19.8  & 19.1  & 18.4 \\[0.5mm]\hline
 $f_+(q^2)$ & $9.04$ & $6.32$ & $4.75$ & $3.75$ & $3.06$ & $2.56$ & $2.19$ & $1.91$ & $1.69$ & $1.51$ & $1.37$ & $1.27$ \\[0.5mm]\hline
statistics + $\chi$PT  (\%) & 24.4  & 18.5  & 13.5  & 9.6  & 7.1  & 6.3  & 6.5  & 6.9
  & 7.2  & 7.5  & 8.2  & 9.8 \\[0.5mm]\hline
$g_{B^* B \pi}$ uncertainty  & 1.1  & 0.3  & 0.8  & 1.8  & 2.4  & 2.8  & 2.9  & 2.8  & 2.6  & 2.5  & 2.6  & 2.9 \\
$r_1$  & 0.4  & 0.7  & 0.9  & 1.1  & 1.2  & 1.3  & 1.4  & 1.4  & 1.5 & 1.5  & 1.4  & 1.4 \\
$\hat{m}$  & 0.2  & 0.2  & 0.2  & 0.2  & 0.2  & 0.2  & 0.2  & 0.3  & 0.3  & 0.3  & 0.3  & 0.3 \\
$m_{s}$ & 0.6  & 0.6  & 0.6  & 0.7  & 0.7  & 0.8  & 0.8  & 0.9  & 1.0  & 1.1  & 1.2  & 1.3 \\
$m_b$ & 1.2 & 1.2 & 1.2 & 1.2 & 1.2 & 1.2 & 1.2 & 1.2 & 1.2 & 1.2 & 1.2 & 1.2 \\
heavy quark discretization & 3.4 & 3.4 & 3.4 & 3.4 & 3.4 & 3.4 & 3.4 & 3.4 & 3.4 & 3.4 & 3.4 & 3.4 \\
nonperturbative $Z_V$ & 1.4  & 1.4  & 1.4  & 1.4  & 1.4  & 1.4  & 1.4  & 1.4  & 1.4  & 1.4  & 1.4  & 1.4 \\
perturbative $\rho$ & 3.0  & 3.0  & 3.0  & 3.0  & 3.0  & 3.0  & 3.0  & 3.0  & 3.0  & 3.0  & 3.0  & 3.0 \\
$u_0$ & 2.9  & 2.1  & 1.2  & 0.5  & 0.3  & 0.8  & 1.1  & 1.3  & 1.4  & 1.3  & 1.3  & 1.3 \\
finite volume & 0.5  & 0.5  & 0.5  & 0.5  & 0.5  & 0.5  & 0.5  & 0.5  & 0.5  & 0.5  & 0.5  & 0.5 \\
\hline
total systematics (\%) & 5.9  & 5.4  & 5.3  & 5.4  & 5.7  & 5.9  & 6.0  & 6.1  & 6.0  & 6.0  & 6.0  & 6.2 \\[0.5mm]\hline\hline
\end{tabular}\end{sidewaystable}

\section{\boldmath Model-independent determination of $|V_{ub}|$}
\label{sec:Vub}

It is well-established that analyticity, crossing symmetry, and unitarity largely constrain the possible shapes of semileptonic form factors~\cite{Bourrely:1980gp,Boyd:1994tt,Lellouch:1995yv,Boyd:1997qw}.  In this section we apply constraints based on these general properties to our lattice result for the form factor $f_+(q^2)$ and thereby extract a model-independent value for the CKM matrix element $|V_{ub}|$.  

\bigskip

Until now the standard procedure used to extract $|V_{ub}|$ from $B\to\pi\ell\nu$ semileptonic decays has been to integrate the form factor $|f_+(q^2)|^2$ over a region of $q^2$, and then combine the result with the experimentally measured decay rate in this region:
\begin{equation}
	\frac{\Gamma (q_\textrm{min})}{|V_{ub}|^2} = \frac{G_F^2}{192 \pi^3 m_B^3}  \int_{q^2_\textrm{min}}^{q^2_\textrm{max}} dq^2 \left[ (m_B^2 + m_\pi^2 - q^2)^2 - 4 m_B^2 m_\pi^2 \right]^{3/2} |f_+(q^2)|^2 .
\label{eq:Vub_standard}
\end{equation}
The integration, however, necessitates a continuous parameterization of the form factor over the full range from $q^2_\textrm{min}$ to $q^2_\textrm{max}$.  

In our earlier, preliminary unquenched analysis, we determine $f_+(q^2)$ by fitting the lattice data points to the Be\'cirevi\'c-Kaidalov (BK) parameterization~\cite{Becirevic:1999kt},
\begin{eqnarray}
	f_+(q^2) & = &  \frac{f_+(0)}{\left(1-\tilde{q}^2\right) \left(1-\alpha \, \tilde{q}^2\right)}, \label{eq:BK_f+}\\
	f_0(q^2) & = & \frac{f_+(0)}{\left(1-\tilde{q}^2/\beta \right)} ,
\end{eqnarray}
where $\tilde{q}^2 \equiv q^2/m_{B^*}^2$.  The BK ansatz contains three free parameters and incorporates many of the known properties of the form factor such as the kinematic constraint at $q^2=0$, heavy-quark scaling, and the location of the $B^*$ pole.  The HPQCD Collaboration instead uses the four-parameter Ball-Zwicky (BZ) parameterization~\cite{Ball:2004ye}, which is the same as the BK function in Eq.~(\ref{eq:BK_f+}) plus an additional pole to capture the effects of multiparticle states. 
In both cases, however, the choice of fit function introduces a systematic uncertainty that is difficult to quantify.

It is likely the BK and BZ parameterizations can be safely used to interpolate between data points, whether they be at high $q^2$ from lattice QCD or at low $q^2$ from experiment.
It is less clear, however, how well these ansatze can be trusted to extrapolate the form factor shape beyond the reach of the data points.
Furthermore, comparisons of lattice and experimental determinations of BK or BZ fit parameters are not necessarily meaningful.
For example, if the slope parameters $\alpha$ from experiment and lattice QCD were found to be inconsistent, we would not know whether theory and experiment disagree, or whether the parameterization is simply inadequate.
A parameterization that circumvents this issue is therefore desirable.  In this work we pursue an analysis based on the model-independent $z$-parameterization, which is pedagogically reviewed in Ref.~\cite{Boyd:1994tt}.

\subsection{Analyticity, unitarity, and heavy-quark constraints on heavy-light form factors}

All form factors are analytic functions of $q^2$ except at physical poles and threshold branch points.  In the case of the $B\to \pi l \nu$ form factors,  $f(q^2)$ is analytic below the $B\pi$ production region except at the location of the $B^*$ pole.  The fact that analytic functions can always be expressed as convergent power series allows the form factors to be written in a particularly useful manner.   

Consider mapping the variable $q^2$ onto a new variable, $z$, in the following way:
\begin{equation}
z(q^2, t_0) = \frac{\sqrt{1 - q^2/t_+}-\sqrt{1-t_0/t_+}}{\sqrt{1-q^2/t_+}+\sqrt{1-t_0/t_+}} ,
\label{eq:zee_var}
\end{equation}
where $t_+\equiv(m_B + m_\pi )^2$, $t_-\equiv(m_B - m_\pi )^2$, and $t_0$ is a free parameter.  Although this mapping appears complicated, it actually has a simple interpretation in terms of $q^2$;  this transformation maps $q^2 > t_+$ (the production region) onto $|z|=1$  and maps $q^2 < t_+$ (which includes the semileptonic region) onto real $z \in [-1,1]$.   In terms of $z$, the form factors have a simple form:
\begin{equation}
f(q^2) = \frac{1}{P(q^2) \phi(q^2,t_0)}  \sum_{k=0}^{\infty} a_k(t_0) z(q^2,t_0)^k ,
\label{eq:z_exp}
\end{equation}
where the Blaschke factor $P(q^2)$ is a function that contains subthreshold poles and the outer function $\phi(q^2,t_0)$ is an arbitrary analytic function (outside the cut from $ t_+ < q^2 < \infty$) whose choice only affects the particular values of the series coefficients $a_k$.  

For the case of the $B\to\pi\ell\nu$ form factor $f_+(q^2)$, the Blaschke factor $P_+(q^2) = z(q^2, m_{B^*}^2)$  accounts for the $B^*$ pole.  In this work we use the same outer function as in Ref.~\cite{Arnesen:2005ez}:
\begin{eqnarray}
	\phi_+(q^2, t_0) & = & \sqrt{\frac{3}{96 \pi \chi^{(0)}_J}} \left( \sqrt{t_+ - q^2} + \sqrt{t_+ - t_0}  \right) \left( \sqrt{t_+ - q^2} + \sqrt{t_+ - t_-} \right)^{3/2}  \nonumber\\
				& \times & \left(  \sqrt{t_+ - q^2} + \sqrt{t_+}  \right)^{-5} \frac{(t_+ - q^2)}{(t_+ - t_0)^{1/4}} ,
\label{eq:phi}\end{eqnarray}
where $\chi^{(0)}_J$ is a numerical factor that can be calculated via the operator product expansion (OPE)~\cite{Lellouch:1995yv,Generalis:1990id}.  This choice of  $\phi_+(q^2, t_0)$, when combined with unitarity and crossing-symmetry, leads to a particularly simple constraint on the series coefficients in Eq.~(\ref{eq:z_exp}).  Although the $t$-dependence of Eq.~(\ref{eq:phi}) appears complicated, it is designed so that the sum over the squares of the series coefficients is $t$-independent:
\begin{equation}
	\sum_{k=0}^{\infty} a_k^2 = \frac{1}{2 \pi i} \oint \frac{dz}{z} | P(z) \phi(z) f(z) |^2 \equiv A,
\label{eq:ff_int}
\end{equation}
where the value of the constant $A$ depends upon the choice of $\chi^{(0)}_J$ in Eq.~(\ref{eq:phi}).  Because the decay process $B\to\pi\ell\nu$ is related to the scattering process $\ell\nu\to B \pi$ by crossing symmetry, the sum of the series coefficients is bounded by unitarity, \emph{i.e.}, the fact that the production rate of $B\pi$ states is less than or equal to the production of all final states that couple to the $b\to u$ vector current.  In particular, if one chooses the numerical factor $\chi^{(0)}_J$ to be equal to the appropriate integral of the inclusive rate $\ell\nu \to X_b$, the sum of the coefficients is bounded by unity:
\begin{equation}
\sum_{k=0}^{N} a_k^2 \ltapprox 1,
\label{eq:a_const}
\end{equation}
where this constraint holds for any value of  $N$ and the ``$\ltapprox$" symbol indicates higher-order corrections to $\chi^{(0)}_J$ in $\alpha_s$ and the OPE.  

Such higher-order corrections turn out to be negligible for the $B\to\pi\ell\nu$ form factor because the bound in Eq.~(\ref{eq:a_const}) is far from saturated, \emph{i.e.}, the sizes of the coefficients turn out to be much less than one.  Becher and Hill~\cite{Becher:2005bg} have pointed out that this is due to the fact that the $b$-quark mass is so large.  In the heavy-quark limit, the leading contributions to the integral in Eq.~(\ref{eq:ff_int}) are of $\CO(\Lambda^3 / m_b^3)$, where $\Lambda$ is a typical hadronic scale.  Assuming that the ratio $\Lambda / m_b \sim 0.1$, the heavy-quark bound on the $a_k$'s is approximately thirty times more constraining than the bound from unitarity alone:
\begin{equation}
	\sum_{k=0}^{N} a_k^2 \sim \left( \frac{\Lambda}{m_B}\right)^3 \approx 0.001 .
\label{eq:hq_const}\end{equation}

We point out that the authors of Ref.~\cite{Bourrely:2008za} have recently proposed a slightly different parameterization of the $B\to\pi\ell\nu$ form factor with a simpler choice of outer function, $\phi = 1$:
\begin{equation}
f_+(q^2) = \frac{1}{1-q^2/m_{B^*}^2}  \sum_{k=0}^{\infty} b_k(t_0) z(q^2,t_0)^k .
\end{equation}
This choice enforces the correct scaling behavior, $f_+(q^2) \sim 1/q^2$ as $q^2 \to \infty$.  It leads, however, to a more complicated constraint on the series coefficients:
\begin{equation}
\sum_{j,k=0}^{N} B_{jk} b_j b_k \ltapprox 1,
\end{equation}
where the elements of the symmetric matrix $B_{jk}$ are calculable functions of $t_0$.  Because $B\to\pi\ell\nu$ semileptonic decay is far from $q^2 \to \infty$, and because the unitary bound is so far from being saturated, the choice of outer function should make a negligible impact on the resulting determination of $|V_{ub}|$.  We therefore use the more standard outer function given in Eq.~(\ref{eq:phi}) because the constraint in Eq.~(\ref{eq:a_const}) is independent of the number of terms in the power series, and is therefore simpler to implement.

The free parameter $t_0$ can be chosen to make the maximum value of $|z|$ as small as possible in the semileptonic region;  we choose  $t_0=0.65t_-$  as in Ref.~\cite{Arnesen:2005ez}.  For $B\to \pi l \nu$ semileptonic decays this maps the physical region onto:
\begin{equation}
	0 < t < t_- \;\;  \mapsto \;\; -0.34 < z < 0.22 .
\end{equation}
The bound on the coefficients in the $z$-expansion combined with the small numerical values of  $|z|$ in the physical region ensures that one needs only the first few terms in the $z$-expansion to accurately describe the form factor shape.  Moreover, as the precision of both the lattice calculations and experimental measurements improve, one may easily include higher-order terms as needed.  

\subsection{\boldmath Determination of $|V_{ub}|$ using $z$-parameterization}

In 2007 the BABAR Collaboration published a measurement of the shape of the $B\to\pi\ell\nu$ semileptonic form factor with results for 12 separate $q^2$ bins between $q^2_\textrm{min} \approx 1 \textrm{ GeV}^2$ and $q^2_\textrm{max} \approx 24 \textrm{ GeV}^2$~\cite{Aubert:2006px}.  This suggests that lattice QCD calculations are now needed primarily to provide a precise form factor normalization at one value of $q^2$ in order to determine $|V_{ub}|$.  The minimal error in $|V_{ub}|$ can, of course, still be attained by using all of the available information on the form factor shape and normalization, provided that one analyzes the data in a model-independent way.  

Because as many terms can be added as are needed to describe the $B\to\pi\ell\nu$ form factor to the desired accuracy, use of the convergent series expansion allows for a systematically improvable determination of $|V_{ub}|$.  We fit our lattice numerical Monte Carlo data and the 12-bin BABAR experimental data together to the $z$-expansion, leaving the relative normalization factor, $|V_{ub}|$, as a free parameter to be determined by the fit.  In this way we determine $|V_{ub}|$ in an optimal, model-independent way. 

We first fit the lattice numerical Monte Carlo data and the 12-bin BABAR experimental data \emph{separately} to the $z$-expansion in order to check for consistency.   We use Gaussian priors with central value 0 and width 1 on each coefficient in the $z$-expansion to impose the unitarity constraint.  Although this manner of constraining the coefficients is less stringent than the strict bound given in Eq.~(\ref{eq:a_const}), the choice does not matter because the unitary bound is far from saturated and the individual coefficients all turn out to be much less than 1.  We obtain identical fit results even when the coefficients are completely unconstrained.

The left-hand plot in Fig.~\ref{fig:BABAR_ff_data} shows the BABAR measurement of the $B\to\pi\ell\nu$ semileptonic form factor, $f_+(q^2)$~\cite{Aubert:2006px}.  The right-hand plot shows the same data multiplied by the functions $P_+(q^2)$ and $\phi_+(q^2, t_0)$ and plotted versus the variable $z$.  After remapping from $q^2$ to $z$ there is almost no curvature in the experimental data.  This indicates that most of the curvature in the data is due to well-understood QCD effects that are parameterized by the functions $P_+(q^2)$ and $\phi_+(q^2, t_0)$.  Consequently the experimental data is well-described by a normalization ($a_0$) and slope ($a_1/a_0$), as shown in Fig.~\ref{fig:BABAR_ff_data}.  The slope of the BABAR experimental $B\to\pi\ell\nu$ form factor data is
\begin{equation}
	\frac{a_1}{a_0} = - 1.60 \pm 0.26.
\end{equation}
If one includes a curvature term in the $z$-fit, the coefficient $a_2$ is poorly determined, but is found to be negative at $\sim 1.5 \sigma$.  The value of $a_1$ is consistent with the result of the linear fit.

\begin{CFigure}
	\begin{tabular}{cc}
	\hspace{-0.00 \textwidth}
	\rotatebox{0}{\includegraphics[width=0.47\textwidth]{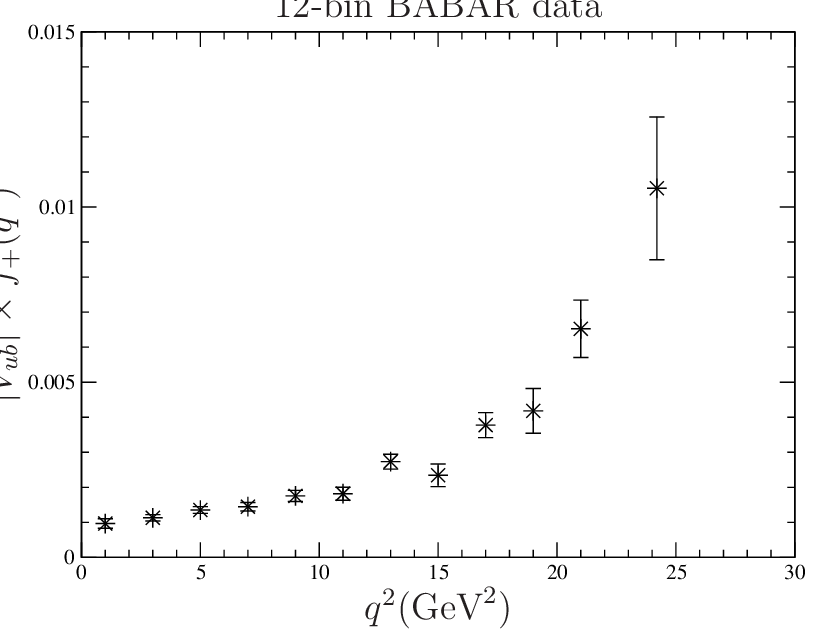}}
	& \hspace{0.02 \textwidth} 
	\rotatebox{0}{\includegraphics[width=0.47\textwidth]{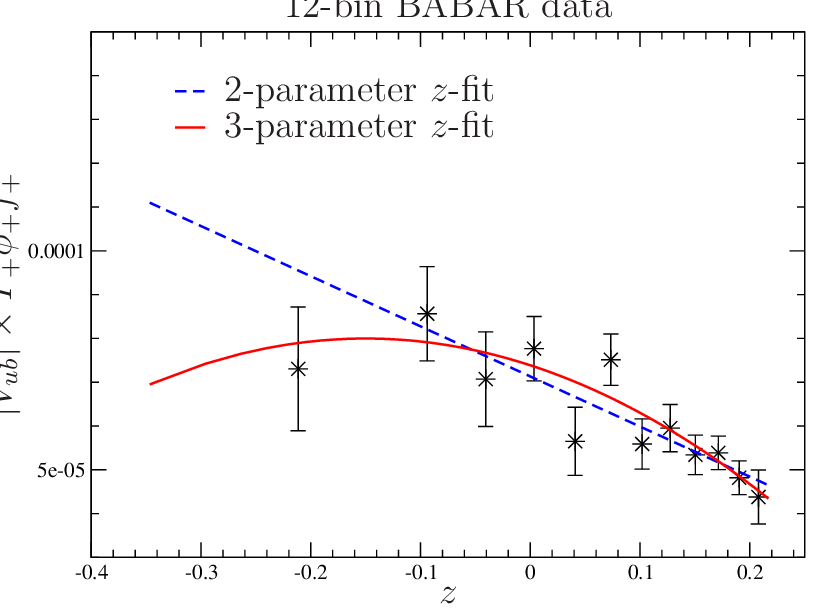}} 
	\end{tabular}
	\caption{Experimental data for the $B\to \pi l \nu$ form factor times the CKM element $|V_{ub}|$ from the BABAR collaboration~\cite{Aubert:2006px}.  The left plot shows $|V_{ub}| \times f_+$ versus $q^2$ while the right plot shows $|V_{ub}| \times f_+$ multiplied by the functions $P_+ \phi_+$  and plotted against  the new variable $z$.  Both the 2-parameter fit (dashed blue line) and 3-parameter fit (solid red curve) have good $\chi^2/\text{d.o.f.}$'s.}
	\label{fig:BABAR_ff_data}
\end{CFigure}

Figure~\ref{fig:lat_ff_data} shows the lattice determination of the $B\to\pi\ell\nu$ semileptonic form factor, $f_+$ vs. $q^2$ (left plot) and the remapped form factor, $P_+ \phi_+ f_+$ vs. $z$ (right plot).  As is the case for the experimental data, the shape of the lattice form factor is less striking after taking out the $B^*$ pole and other known QCD effects.  When the lattice calculation of the form factor is fit to the $z$-parameterization, however, it determines both a slope and a curvature.  One cannot, in fact, successfully fit the lattice data without including a curvature term.  The slope and curvature of the lattice determination of the $B\to\pi\ell\nu$ form factor are
\begin{eqnarray}
	\frac{a_1}{a_0} &=& -1.75 \pm 0.91,  \\
	\frac{a_2}{a_0} &=& -5.22 \pm 1.39 . 
\end{eqnarray}
The above uncertainties are the standard errors computed from the inverse of the parameter Hessian matrix that result from a fit using the full covariance matrix determined from the bootstrap distributions of chiral-continuum extrapolated values of $f_\parallel$ and $f_\perp$, including systematics.

\begin{CFigure}
	\begin{tabular}{cc}
	\hspace{-0.00 \textwidth}
	\rotatebox{0}{\includegraphics[width=0.469\textwidth]{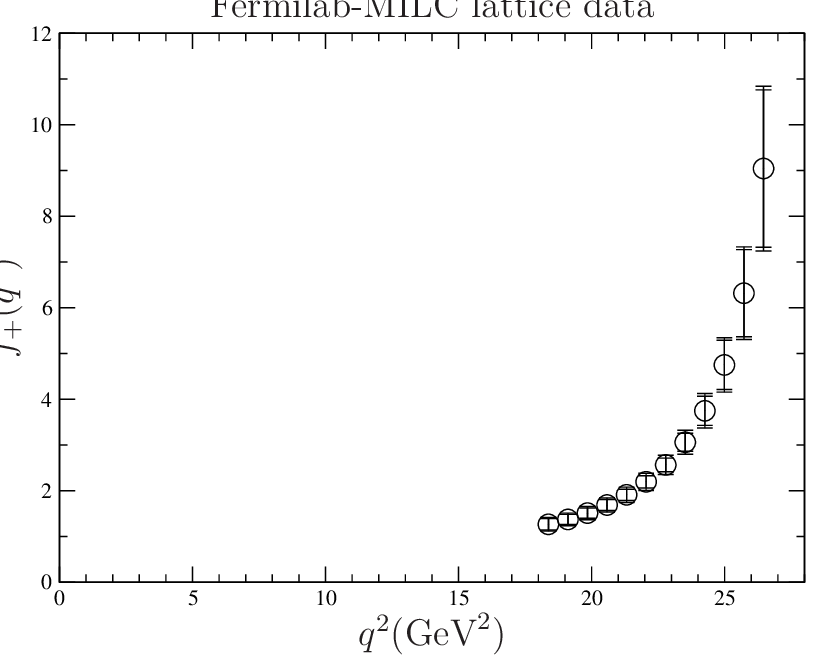}}
	& \hspace{0.035 \textwidth}
	\rotatebox{0}{\includegraphics[width=0.469\textwidth]{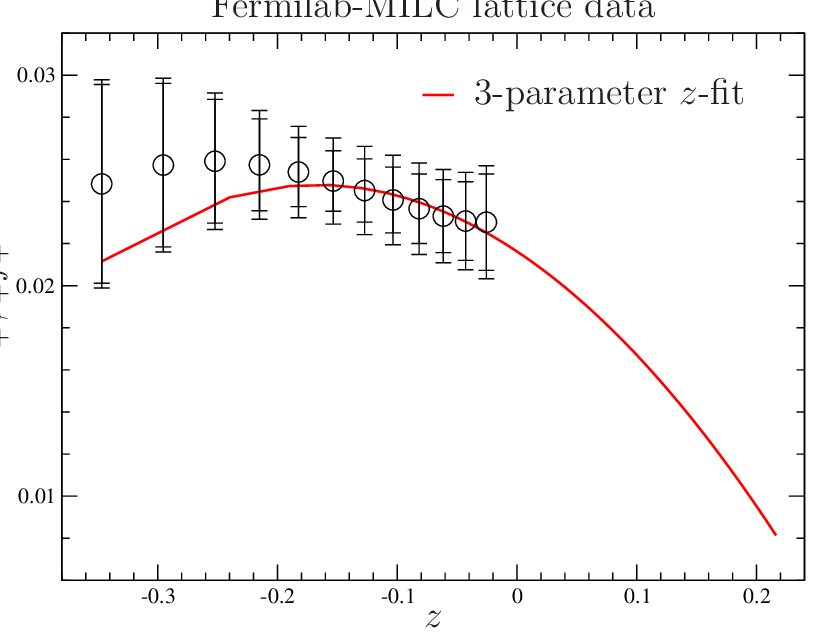}} 
	\end{tabular}
	\caption{Lattice calculation of the $B\to \pi l \nu$ form factor.  The left plot shows $f_+$ vs. $q^2$ while the right plot shows $P_+ \phi_+ f_+$ vs. $z$.  The inner error bars indicate the statistical error, while the outer error bars indicate the sum of the statistical and systematic added in quadrature.  A 3-parameter $z$-fit is needed to describe the lattice data with a good $\chi^2/$d.o.f.}
	\label{fig:lat_ff_data}
\end{CFigure}

Because the shapes of the lattice calculation and experimental measurement of the form factor are consistent, we now proceed to fit them simultaneously to the $z$-expansion and determine $|V_{ub}|$.  The numerical lattice and measured experimental data are independent, so we construct a block-diagonal covariance matrix where one block is the total lattice error matrix and the other is the total experimental error matrix.  The combined fit function includes the series coefficients ($a_k$'s) plus an additional parameter for the relative normalization between the lattice and experimental results ($|V_{ub}|$).  In order to account for the systematic uncertainty in $|V_{ub}|$ due to poorly-constrained higher-order terms in $z$, we continue to add terms in the series until the error in $|V_{ub}|$ reaches a maximum.  This occurs once we include the term proportional to $z^3$.  The resulting combined $z$-fit is shown in Fig.~\ref{fig:sim_zfit}, and the corresponding fit parameters are
\begin{eqnarray}
	|V_{ub}| \times 10^3 &=& 3.38 \pm 0.36 \,, \label{eq:Vub_res}\\
	a_0 &=& 0.0218 \pm  0.0021 \,, \\
	a_1 &=& -0.0301 \pm  0.0063 \,, \\
	a_2 &=& -0.059 \pm   0.032 \,, \\
	a_3 &=& 0.079 \pm  0.068 \,.
\end{eqnarray}
The values of the coefficients are all much smaller than one, as expected from heavy-quark power-counting.  The sum of the squares of the coefficients is $\sum a_k^2 = 0.011 \pm 0.012$, and is consistent with the prediction of Becher and Hill within uncertainties in the series coefficients and in the choice of the hadronic scale in Eq.~(\ref{eq:hq_const})~\cite{Becher:2005bg}.

\begin{CFigure}
	\begin{tabular}{cc}
	\hspace{0.015 \textwidth}
	\rotatebox{0}{\includegraphics[width=0.47\textwidth]{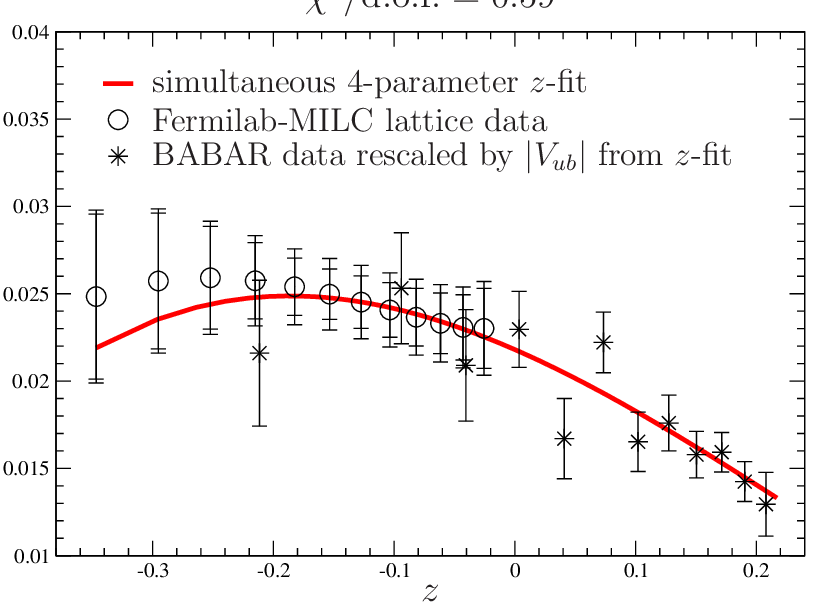}}
	& \hspace{0.015 \textwidth}
	\rotatebox{0}{\includegraphics[width=0.47\textwidth]{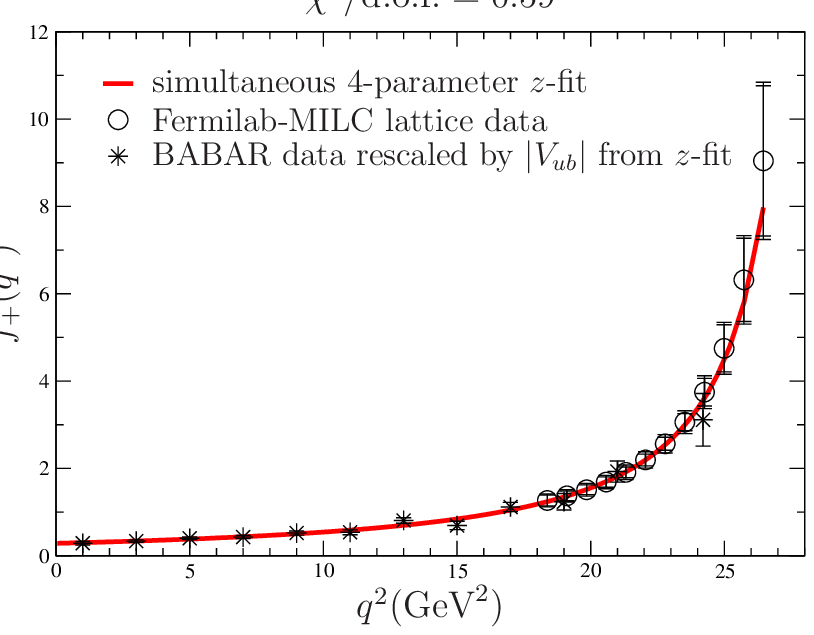}}
	\end{tabular}
	\caption{Model-independent determination of $|V_{ub}|$ from a simultaneous fit of lattice and experimental $B\to\pi\ell\nu$ semileptonic form factor data to the $z$-parameterizaton.  The left plot shows $P_+ \phi_+ f_+$ vs. $z$ while the right plot shows $f_+$ vs. $q^2$.  Inclusion of terms in the power-series through $z^3$ yields the maximum uncertainty in $|V_{ub}|$;  the corresponding 4-parameter $z$-fit is given by the red curve in both plots. The circles denote the Fermilab-MILC lattice data, while the stars indicate the 12-bin BABAR experimental data, rescaled by the value of $|V_{ub}|$ determined in the simultaneous $z$-fit.}
	\label{fig:sim_zfit}
\end{CFigure}

By combining all of the available numerical lattice Monte Carlo data and 12-bin BABAR experimental data for the $B\to\pi\ell\nu$ form factor in a simultaneous fit we are able to determine $|V_{ub}|$ to $\sim 11\%$ accuracy.  This error is independent (within $\ltapprox 0.5\%$) of the choice of the parameter $t_0$ used in the change of variables from $q^2$ to $z(q^2, t_0)$ and in the outer function $\phi_+(q^2, t_0)$.  In order to demonstrate the advantage of the combined fit method, we compare the error in $|V_{ub}|$ given in Eq.~(\ref{eq:Vub_res}) with that obtained from separate $z$-fits of the lattice and experimental data.  A $z$-fit to the 12-bin BABAR experimental data alone determines the normalization ${a_0}^\text{exp}$ to $\sim 8\%$, while a $z$-fit to our numerical lattice data determines ${a_0}^\text{lat}$ to $\sim 14\%$.  Thus separate fits lead to a determination of $|V_{ub}| \equiv {a_0}^\text{exp} / {a_0}^\text{lat}$ with an approximately $16 \%$ total uncertainty.\footnote{Because the values of the coefficients of the power-series in $z$ depend upon the choice of the parameter $t_0$ in Eqs.~(\ref{eq:zee_var})--(\ref{eq:phi}), we could, in principle, choose a different value of $t_0$ in order to minimize the error in either ${a_0}^\text{exp}$ or ${a_0}^\text{lat}$.  For example, use of $t_0=22.8 \text{ GeV}^2$ reduces the uncertainty in the lattice normalization because the error in the lattice form factor is smallest at this $q^2$-value.  Use of $t_0=22.8 \text{ GeV}^2$ greatly increases the uncertainty in the experimental normalization, however, because the experimental data is poorly-determined at large values of $q^2$.  Ultimately, this choice of $t_0$ leads to an even worse determination of $|V_{ub}|$ than from our standard choice of $t_0 = 0.65 t_-$.  Although we did not attempt to determine the value of $t_0$ that minimizes the total error in $|V_{ub}|$, the errors resulting from separate fits were greater than that obtained with the simultaneous fit for all values of $t_0$ that we tried.}  The combined fit yields a significantly smaller error and is thus preferred.

When the numerical lattice data and experimental data are fit simultaneously, utilizing all of the available data points is of secondary importance for reducing the total uncertainty in $|V_{ub}|$.  For example, we can evaluate the importance of the low $q^2$ experimental points to the extraction of $|V_{ub}|$ by removing them from the combined $z$-fit.  Including only the three experimental data points with $q^2 > 18$ GeV$^2$, we find a consistent value of $|V_{ub}|$ with only a $\sim 1\%$ larger uncertainty.  Similarly, we can evaluate the importance of having many lattice data points, rather than only a single point, by using only the most precise lattice point with a total error of $\sim 9 \%$.  This allows the form factor shape to be completely determined by the experimental data.   We find a consistent value of $|V_{ub}|$ but with an even larger error of $\sim 13\%$.  We therefore conclude that combining all of the numerical lattice data with all of the experimentally-measured BABAR data minimizes the total uncertainty in $|V_{ub}|$.  Because the small error in our final determination of $|V_{ub}|$ is primarily due to the power of the combined $z$-fit method, one could easily use the procedure outlined in this section to improve the exclusive determination of $|V_{ub}|$ from existing lattice QCD calculations of the $B\to\pi\ell\nu$ form factor such as that by the HPQCD Collaboration~\cite{Dalgic:2006dt}.

\section{Results and Conclusions}
\label{sec:Conc}

Combining our latest unquenched lattice calculation of the $B\to\pi\ell\nu$ form factor with the 12-bin BABAR experimental data, we find the following model-independent value for $|V_{ub}|$:\footnote{At three conferences during Summer 2008 we presented a version of this model-independent analysis with a numerical value for $|V_{ub}|$ that is 1-$\sigma$ lower than that given here in Eq. (79).  We have since improved several aspects of the lattice calculation, most notably reducing the statistical errors that enter the chiral and continuum extrapolations of $f_\perp$ and~$f_\parallel$ and, hence, $f_+$.  Equation~(\ref{eq:answer}) is our final result for $|V_{ub}|$ based on the lattice data from the ensembles in Table~\ref{tab:MILC_ens} and the methodology of Secs.~\ref{sec:Analysis} and~\ref{sec:Vub}.}
\begin{equation}
\label{eq:answer}
	|V_{ub}| \times 10^3= 3.38 \pm 0.36 .
\end{equation}
The total error is $\sim 11 \%$, and it is nontrivial to separate the error precisely into contributions from statistical, systematic, and experimental uncertainty because of the combined $z$-fit procedure used.  If we assume, however, that the error in $|V_{ub}|$ is dominated by the most precisely determined lattice point (which is not quite true, as shown in the previous section), we can estimate that the contributions are roughly equally divided as $\sim 6\%$ lattice statistical, $\sim 6\%$ lattice systematic, and $\sim 6\%$ experimental.

Our result is consistent with, although slightly lower than, our earlier, preliminary determination of $|V_{ub}|$.  The reduction in central value is primarily due to a change in the lattice determination of the form factor, not the procedure used to determine $|V_{ub}|$.  Because our new analysis uses a second lattice spacing, we are able to take the continuum limit of the form factor.  We find that the continuum extrapolation increases the overall normalization of $f_+(q^2)$, and hence decreases the value of $|V_{ub}|$.  Our errors are smaller than those of previous exclusive determinations primarily because we have reduced the size of the discretization errors, which are significantly smaller than in the previous Fermilab-MILC calculation ($\sim 7\% \to 3\%$) because of the additional finer lattice spacing.
  
Our new result is $\sim 1$--$2 \sigma$ lower than most inclusive determinations of $|V_{ub}|$, which typically range from $4.0 - 4.5 \times 10^{-3}$~\cite{HFAG_ICHEP08}.   Much of the variation among the inclusive values is due to the choice of input parameters --- in particular that of the $b$-quark mass~\cite{Gambino_CKM08}.  The recent determination of $m_b$ by K\"uhn, Steinhauser, and Sturm using experimental data for the cross section for $e^+ e^- \to$ hadrons in the bottom threshold region yields the value of $m_b$ to percent-level accuracy~\cite{Kuhn:2007vp}, and is consistent with the PDG average~\cite{Amsler:2008zz}.  Neubert has shown, however, that an updated extraction of $m_b$ from fits to $B \to X_c \ell \nu$ moments using only the theoretically cleanest channels (excluding $b\to X_s \gamma$) results in a larger $b$-quark mass and hence smaller inclusive value of $|V_{ub}|$, thereby reducing the tension between inclusive and exclusive determinations~\cite{Neubert:2008cp}. 
 
Our result is consistent with the currently preferred values for $|V_{ub}|$ determined by the global CKM unitarity triangle analyses of the CKMfitter Collaboration, $|V_{ub}| \times 10^3 = 3.44^{+0.22}_{-0.17}$~\cite{CKMfitter}, and UTFit Collaboration, $|V_{ub}| \times 10^3 = 3.48 \pm 0.16$,~\cite{UTFit}.  Further reduction in the errors is therefore essential for a more stringent test of the CKM framework and a more sensitive probe of physics beyond the Standard Model.
 
The dominant uncertainty in our lattice calculation of the $B\to\pi\ell\nu$ form factor comes from the statistical errors in the 2-point and 3-point correlations.  This error can be reduced in a straightforward manner with use of an improved source for the pion and/or additional gauge configurations.  The statistical errors in the nonperturbative renormalization factors $Z_V^{bb}$ and $Z_V^{ll}$ can be brought to below a percent in the same way.  The chiral-continuum extrapolation error, which is inextricably linked to the statistical errors in the correlation functions, can also be improved by simulating at more light quark masses and an additional finer lattice spacing of $a \sim 0.06$ fm.  Presumably a better constrained chiral and continuum extrapolation will reduce the size of other $q^2$-dependent errors such as those from $g_{B^* B \pi}$, $r_1$, and the light quark masses by some unknown amount as well.   
Use of a finer lattice with $a \sim 0.06$ fm will further decrease the momentum-dependent and heavy-quark discretization errors, which we now estimate with power-counting.  The extraction of $|V_{ub}|$ can also be improved by including more experimental measurements of the $B\to\pi\ell\nu$ branching fraction.  This, however, will require understanding the correlations among the various systematic uncertainties.  Given these refinements of the current calculation, an even more precise, model-independent value of $|V_{ub}|$ can be obtained in the near future.  

\section*{Acknowledgments}

We thank Richard Hill and Enrico Lunghi for helpful discussions, and Thomas Becher for valuable comments on the manuscript.
Computations for this work were carried out in part on facilities of
the USQCD Collaboration, which are funded by the Office of Science of
the U.S. Department of Energy, and on facilities of the NSF Teragrid under allocation TG-MCA93S002.
This work was supported in part by the United States Department of Energy
under Grant Nos.~DE--FC02-06ER41446 (C.D., L.L.), DE-FG02-91ER40661(S.G.), DE-FG02-91ER40677 (A.X.K., R.T.E., E.G.), DE-FG02-91ER40628  (C.B., J.L.), DE-FG02-04ER41298 (D.T.) and by the National Science Foundation under Grant Nos.~PHY-0555243, PHY-0757333, PHY-0703296 (C.D., L.L.), PHY-0555235 (J.L.), PHY-0456556 (R.S.).  R.T.E. and E.G. were supported in part by URA visiting scholar awards.
Fermilab is operated by Fermi Research Alliance, LLC, under Contract 
No.~DE-AC02-07CH11359 with the United States Department of Energy.

\appendix
\section{Estimate of heavy quark discretization errors}
\label{sec:App}

In this Appendix we collect the short-distance functions~$f_i$ used to 
estimate the heavy-quark discretization effects.
For more background, see Refs.~\cite{ElKhadra:1996mp,Kronfeld:2000ck,%
Harada:2001fi,Harada:2001fj,Kronfeld:2003sd,Oktay:2008ex}.

\subsection{\boldmath $\CO(a^2)$ errors}

We start with these because explicit expressions for the functions $f_i(m_0a)$ are available.

\subsubsection{$\CO(a^2)$ errors from the Lagrangian}

There are two bilinears, $\bar{h}\vec{D}\cdot\vec{E}h$ and
$\bar{h}i\vec{\Sigma}\cdot[\vec{D}\times\vec{E}]h$, and many four-quark operators.
At tree level the coefficients of all four-quark operators vanish and the coefficients of the two bilinears are the same.  The mismatch function is given by
\begin{equation}
	f_E(m_0a) = \frac{1}{8m_E^2a^2} - \frac{1}{2(2m_2a)^2}.
\end{equation}
Using explicit expressions for $1/m_2$~\cite{ElKhadra:1996mp} and $1/m_E^2$~\cite{Oktay:2008ex}, one finds
\begin{equation}
	f_E(m_0a) = \frac{1}{2}\left[
		\frac{c_E(1+m_0a)-1}{m_0a(2+m_0a)(1+m_0a)} -
		\frac{1}{4(1+m_0a)^2} \right].
\end{equation}
We use $c_E=1$ in our numerical simulations.

\subsubsection{$\CO(a^2)$ errors from the current}

There are three terms with non-zero coefficients,
$\bar{q}\Gamma\vec{D}^2h$,
$\bar{q}\Gamma i\vec{\Sigma}\cdot\vec{B}h$, and
$\bar{q}\Gamma\vec{\alpha}\cdot\vec{E}h$, 
which can be deduced from Eq.~(A17) of Ref.~\cite{ElKhadra:1996mp}.
Their coefficients can be read off from Eqs.~(A19)~\cite{ElKhadra:1996mp}.
When $c_B=r_s$ the first two share the same coefficient:
\begin{eqnarray}
	f_X(m_0a) & = & \frac{1}{8m_X^2a^2} -
		\frac{\zeta d_1(1+m_0a)}{m_0a(2+m_0a)} - \frac{1}{2(2m_2a)^2},
		\nonumber \\
		& = & \frac{1}{2}\left[ \frac{1}{(2+m_0a)(1+m_0a)} +
			\frac{1}{2(1+m_0a)} - \frac{1}{4(1+m_0a)^2} - 
			\frac{1}{(2+m_0a)^2} \right], \nonumber \\
	& = & \frac{1}{2}\left[ \frac{1}{2(1+m_0a)} - \left(
		\frac{m_0a}{2(2+m_0a)(1+m_0a)} \right)^2\right],
\end{eqnarray}
where the last term on the second line comes from using the 
tree-level~$d_1$.  For the third operator, $\bar{q}\Gamma\vec{\alpha}\cdot\vec{E}h$,
\begin{eqnarray}
	f_Y(m_0a) & = & \frac{1}{2}\left[\frac{d_1}{m_2a} -
		\frac{\zeta(1-c_E)(1+m_0a)}{m_0a(2+m_0a)} \right], \nonumber \\
		& = & \frac{2 + 4 m_0a + (m_0a)^2}{4(1+m_0a)^2(2+m_0a)^2},
\end{eqnarray}
where the last line reflects the choices made for $c_E$ and~$d_1$.

\subsection{\boldmath $\CO(\alpha_s a)$ errors}

Because we improve both the action and current, the mismatch functions $f_i(m_0a)$ start at order $\alpha_s$, and we do not have explicit expressions for them.  (The calculation of these functions would be needed to match at the one-loop level.)  So we shall take unimproved tree-level coefficients as a guide to the combinatoric factors and consider asymptotic behavior in the limits $m_0 a \to 0, \infty$.

\subsubsection{$\CO(\alpha_s a)$ errors from the Lagrangian}

There are two bilinears,
the kinetic energy $\bar{h}\vec{D}^2h$ and the chromomagnetic moment
$\bar{h}i\vec{\Sigma}\cdot\vec{B}h$.
There is no mismatch in the coefficient of the kinetic energy, by 
assumption, since we identify the kinetic mass with the heavy-quark mass.
This tuning is imperfect, but the associated error is budgeted in 
Sec.~\ref{sec:kappab}.

At the tree level the chromomagnetic mismatch is
\begin{equation}
	f^{[0]}_B(m_0a) = \frac{c_B-1}{2(1+m_0a)}.
\end{equation}
This has the right asymptotic behavior in both limits, so our ansatz 
for the one-loop mismatch function is simply
\begin{equation}
	f^{[0]}_B(m_0a) = \frac{\alpha_s}{2(1+m_0a)},
\end{equation}
and ${\tt error}_B$ is this function multiplied by $a\Lambda$.

\subsubsection{$\CO(\alpha_s a)$ errors from the current}

There is only one correction at tree level, but more generally 
there are two for the temporal current and four for the spatial current.
(See Eqs.~(2.27)--(2.32) of Ref.~\cite{Harada:2001fi}.)

The tree-level mismatch function ends up being the same as $d_1$:
\begin{equation}
	f^{[0]}_3(m_0a) = \frac{m_0a}{2(2+m_0a)(1+m_0a)}.
\end{equation}
It is, however, an accident that it vanishes as $m_0a\to0$.
Therefore, we instead take
\begin{equation}
	f_3(m_0a) = \frac{\alpha_s}{2(2+m_0a)},
\end{equation}
which has the right asymptotic behavior.

\subsection{Numerical estimates}

The relative errors due to mismatches in the heavy quark Lagrangian and current on the MILC coarse and fine ensembles are tabulated in Table~\ref{tbl:app:results}.
At the fine lattice spacing we take the typical $\alpha_V(q^*)$ to be
$\frac{1}{3}$, and we use one-loop running to obtain $\alpha_V(q^*)$ at the coarse lattice spacing.    The contribution ${\tt error}_Y$ from the $\vec{\alpha}\cdot\vec{E}$ error in the current is so small both because $c_E=1$ in our simulation and because $d_1$ is small.  Adding the individual errors given in Table~\ref{tbl:app:results} in quadrature, and taking into account multiple contributions of the same size, we find the total error to be 2.84\% (4.16\%) for $f_\parallel$ and 3.40\% (4.98\%) for $f_\perp$ on the fine (coarse) lattices.

\begin{table}[tp]
    \centering
    \caption[tbl:errors]{Relative error from mismatches in the heavy quark
    Lagrangian and current for the bottom quark with $\Lambda=700$~MeV.
    To obtain the totals given in the text $E$ and $X$ are counted
    twice, and $3$ is counted twice for $f_\parallel$ and four times for $f_\perp$.
    Entries are in per cent.}
    \label{tbl:errors}
    \begin{tabular}{cccccccc}
    \hline\hline
    $a$~(fm) & $\alpha_V(q^*)$ & $m_0a$ & $B$ & $3$ & $E$ & $X$ & $Y$ \\\hline
    0.09 & 0.33 & 2.018 & 1.76 & 1.32 & 0.28 & 0.80 & 0.24 \\
    0.12 & 0.41 & 2.617 & 2.48 & 1.94 & 0.39 & 1.26 & 0.33 \\
    \hline\hline
    \end{tabular}
    \label{tbl:app:results}
\end{table}

\newpage
\section{Statistical and systematic error matrices}
\label{sec:Tab}

In this Appendix we present the normalized statistical and systematic bootstrap correlation matrices for the $B\to\pi\ell\nu$ form factor, $f_+(q^2)$, that were used in our model-independent determination of $|V_{ub}|$.  In order to facilitate the use of our result, we also show the resulting total covariance matrix.

\begin{table}
\caption{Normalized statistical (upper) and systematic (lower) bootstrap correlation matrices for the $B\to\pi\ell\nu$ form factor, $f_+(q^2)$.  These should be combined with the values of $f_+(q^2)$ presented in Table~\ref{tab:fP_err} to reconstruct the full correlation matrices. \medskip}
\label{tab:fP_cor}

\begin{tabular}{r|cccccccccccc}
\hline\hline
$q^2$ (GeV$^2$) \ & 26.5  & 25.7  & 25.0  & 24.3  & 23.5  & 22.8  & 22.1  & 21.3  & 20.6  & 19.8  & 19.1  & 18.4 \\[0.5mm]\hline
26.5 \  & 1.00  & 0.99  & 0.97  & 0.88  & 0.66  & 0.29  & -0.01  & -0.16  & -0.19  & -0.14  & -0.04  & 0.05 \\
25.7 \  & 0.99  & 1.00  & 0.99  & 0.92  & 0.73  & 0.38  & 0.07  & -0.09  & -0.13  & -0.09  & 0.00  & 0.08 \\
25.0 \  & 0.97  & 0.99  & 1.00  & 0.97  & 0.82  & 0.51  & 0.21  & 0.04  & -0.02  & 0.01  & 0.07  & 0.13 \\
24.3 \  & 0.88  & 0.92  & 0.97  & 1.00  & 0.93  & 0.69  & 0.42  & 0.25  & 0.18  & 0.17  & 0.20  & 0.21 \\
23.5 \  & 0.66  & 0.73  & 0.82  & 0.93  & 1.00  & 0.91  & 0.72  & 0.56  & 0.48  & 0.43  & 0.39  & 0.32 \\
22.8 \  & 0.29  & 0.38  & 0.51  & 0.69  & 0.91  & 1.00  & 0.94  & 0.85  & 0.77  & 0.69  & 0.58  & 0.42 \\
22.1 \  & -0.01  & 0.07  & 0.21  & 0.42  & 0.72  & 0.94  & 1.00  & 0.98  & 0.92  & 0.84  & 0.69  & 0.48 \\
21.3 \  & -0.16  & -0.09  & 0.04  & 0.25  & 0.56  & 0.85  & 0.98  & 1.00  & 0.98  & 0.92  & 0.78  & 0.55 \\
20.6 \  & -0.19  & -0.13  & -0.02  & 0.18  & 0.48  & 0.77  & 0.92  & 0.98  & 1.00  & 0.97  & 0.86  & 0.66 \\
19.8 \  & -0.14  & -0.09  & 0.01  & 0.17  & 0.43  & 0.69  & 0.84  & 0.92  & 0.97  & 1.00  & 0.95  & 0.80 \\
19.1 \  & -0.04  & 0.00  & 0.07  & 0.20  & 0.39  & 0.58  & 0.69  & 0.78  & 0.86  & 0.95  & 1.00  & 0.94 \\
18.4 \  & 0.05  & 0.08  & 0.13  & 0.21  & 0.32  & 0.42  & 0.48  & 0.55  & 0.66  & 0.80  & 0.94  & 1.00 \\[0.5mm]\hline\hline
\end{tabular}

\bigskip\medskip

\begin{tabular}{r|cccccccccccc}
\hline\hline
$q^2$ (GeV$^2$) \ & 26.5  & 25.7  & 25.0  & 24.3  & 23.5  & 22.8  & 22.1  & 21.3  & 20.6  & 19.8  & 19.1  & 18.4 \\[0.5mm]\hline
26.5 \  & 1.00  & 0.98  & 0.95  & 0.89  & 0.85  & 0.88  & 0.9  & 0.91  & 0.91  & 0.92  & 0.91  & 0.90 \\
25.7 \  & 0.98  & 1.00  & 0.98  & 0.92  & 0.87  & 0.87  & 0.88  & 0.89  & 0.90  & 0.91  & 0.9  & 0.88 \\
25.0 \  & 0.95  & 0.98  & 1.00  & 0.97  & 0.94  & 0.94  & 0.94  & 0.94  & 0.95  & 0.95  & 0.95  & 0.94 \\
24.3 \  & 0.89  & 0.92  & 0.97  & 1.00  & 0.99  & 0.99  & 0.98  & 0.98  & 0.98  & 0.98  & 0.98  & 0.97 \\
23.5 \  & 0.85  & 0.87  & 0.94  & 0.99  & 1.00  & 0.99  & 0.99  & 0.98  & 0.98  & 0.98  & 0.98  & 0.98 \\
22.8 \  & 0.88  & 0.87  & 0.94  & 0.99  & 0.99  & 1.00  & 1.0  & 1.00  & 0.99  & 0.99  & 0.99  & 0.99 \\
22.1 \  & 0.90  & 0.88  & 0.94  & 0.98  & 0.99  & 1.00  & 1.00  & 1.00  & 1.00  & 1.00  & 1.00  & 1.00 \\
21.3 \  & 0.91  & 0.89  & 0.94  & 0.98  & 0.98  & 1.00  & 1.00  & 1.00  & 1.00  & 1.00  & 1.00  & 1.00 \\
20.6 \  & 0.91  & 0.90  & 0.95  & 0.98  & 0.98  & 0.99  & 1.00  & 1.00  & 1.00  & 1.00  & 1.00  & 1.00 \\
19.8 \  & 0.92  & 0.91  & 0.95  & 0.98  & 0.98  & 0.99  & 1.00  & 1.00  & 1.00  & 1.00  & 1.00  & 1.00 \\
19.1 \  & 0.91  & 0.9  & 0.95  & 0.98  & 0.98  & 0.99  & 1.00  & 1.00  & 1.00  & 1.00  & 1.00  & 1.00 \\
18.4 \  & 0.90  & 0.88  & 0.94  & 0.97  & 0.98  & 0.99  & 1.00  & 1.00  & 1.00  & 1.00  & 1.00  & 1.00 \\[0.5mm]\hline\hline
\end{tabular}\end{table}

\begin{table}
\caption{Total bootstrap covariance matrix for the $B\to\pi\ell\nu$ form factor, $f_+(q^2)$, derived by adding the statistical and systematic errors in quadrature. The elements of the matrix are given by $M_{ij} = \sigma_{f_+(q^2_i)} \times \sigma_{f_+(q^2_j)}$, where $\sigma_{f_+(q^2_i)}$ is the total uncertainty in $f_+(q^2)$ in the $i$'th $q^2$ bin. \medskip}
\label{tab:fP_tot_cor}
\begin{tabular}{r|cccccccccccc}
\hline\hline
$q^2$ (GeV$^2$) \ & 26.5  & 25.7  & 25.0  & 24.3  & 23.5  & 22.8  & 22.1  & 21.3  & 20.6  & 19.8  & 
19.1  & 18.4 \\[0.5mm]\hline
26.5 \  & 5.13  & 2.74  & 1.49  & 0.79  & 0.39  & 0.18  & 0.06  & 0.01  & -0.0  & 0.01  & 0.03  & 0.05 \\
25.7 \  & 2.74  & 1.48  & 0.82  & 0.45  & 0.24  & 0.12  & 0.05  & 0.02  & 0.01  & 0.02  & 0.03  & 0.04 \\
25.0 \  & 1.49  & 0.82  & 0.47  & 0.27  & 0.15  & 0.09  & 0.05  & 0.03  & 0.02  & 0.02  & 0.02  & 0.03 \\
24.3 \  & 0.79  & 0.45  & 0.27  & 0.17  & 0.11  & 0.07  & 0.05  & 0.03  & 0.03  & 0.02  & 0.02  & 0.02 \\
23.5 \  & 0.39  & 0.24  & 0.15  & 0.11  & 0.08  & 0.06  & 0.04  & 0.04  & 0.03  & 0.03  & 0.02  & 0.02 \\
22.8 \  & 0.18  & 0.12  & 0.09  & 0.07  & 0.06  & 0.05  & 0.04  & 0.04  & 0.03  & 0.03  & 0.02  & 0.02 \\
22.1 \  & 0.06  & 0.05  & 0.05  & 0.05  & 0.04  & 0.04  & 0.04  & 0.03  & 0.03  & 0.03  & 0.02  & 0.02 \\
21.3 \  & 0.01  & 0.02  & 0.03  & 0.03  & 0.04  & 0.04  & 0.03  & 0.03  & 0.03  & 0.02  & 0.02  & 0.02 \\
20.6 \  & 0.00  & 0.01  & 0.02  & 0.03  & 0.03  & 0.03  & 0.03  & 0.03  & 0.03  & 0.02  & 0.02  & 0.02 \\
19.8 \  & 0.01  & 0.02  & 0.02  & 0.02  & 0.03  & 0.03  & 0.03  & 0.02  & 0.02  & 0.02  & 0.02  & 0.02 \\
19.1 \  & 0.03  & 0.03  & 0.02  & 0.02  & 0.02  & 0.02  & 0.02  & 0.02  & 0.02  & 0.02  & 0.02  & 0.02 \\
18.4 \  & 0.05  & 0.04  & 0.03  & 0.02  & 0.02  & 0.02  & 0.02  & 0.02  & 0.02  & 0.02  & 0.02  & 0.02 \\[0.5mm]\hline\hline
\end{tabular}\end{table}

\newpage


\bibliography{leptonic_08}

\end{document}